\documentclass[conference]{IEEEtran}
\IEEEoverridecommandlockouts
\usepackage{cite}
\usepackage{amsmath,amssymb,amsfonts}
\usepackage{algorithmic}
\usepackage{graphicx}
\usepackage{textcomp}
\usepackage{xcolor}
\def\BibTeX{{\rm B\kern-.05em{\sc i\kern-.025em b}\kern-.08em
    T\kern-.1667em\lower.7ex\hbox{E}\kern-.125emX}}

\usepackage{tcolorbox}
\usepackage{graphicx}
\usepackage{comment}
\usepackage{amsmath}
\usepackage{booktabs}
\usepackage{float}
\usepackage{url}
\usepackage[colorinlistoftodos]{todonotes}

\usepackage{pifont}
\usepackage{graphicx}
\usepackage{multirow}
\usepackage{rotating}
\usepackage{enumitem}
\usepackage{makecell}

\newcommand{\subhead}[1]{\vspace{0.04in}\noindent{\textbf{#1.}}}

\newcommand{\circleTick}{\ding{108}} % Define a filled circle for ticks
\newcommand{\circleNon}{\ding{109}} % Define an empty circle for non    
\begin{document}

\title{Identifying and Addressing User-level Security Concerns in Smart Homes
Using ``Smaller'' LLMs}

\author{
\IEEEauthorblockN{1\textsuperscript{st} Hafijul Hoque Chowdhury\textsuperscript{*}}
\IEEEauthorblockA{ \scriptsize
\textit{Department of CSE} \\
\textit{Bangladesh University of Engineering and Technology} \\
Dhaka, Bangladesh \\
nabid.hasan987@gmail.com}
\and
\IEEEauthorblockN{2\textsuperscript{nd} Riad Ahmed Anonto\textsuperscript{*}}
\IEEEauthorblockA{\scriptsize
\textit{Department of CSE} \\
\textit{Bangladesh University of Engineering and Technology} \\
Dhaka, Bangladesh \\
riadahmedanonto355@gmail.com}
\and
\IEEEauthorblockN{3\textsuperscript{rd} Sourov Jajodia}
\IEEEauthorblockA{\scriptsize
\textit{Concordia Institute for Information Systems Engineering} \\
\textit{Concordia University} \\
Montreal, Canada \\
sourov.jajodia@mail.concordia.ca}
\and
\IEEEauthorblockN{\hspace{1.5cm}4\textsuperscript{th} Suryadipta Majumdar}
\IEEEauthorblockA{\scriptsize
\hspace{1.5cm}\textit{Concordia Institute for Information Systems Engineering} \\
\hspace{1.5cm}\textit{Concordia University} \\
\hspace{1.5cm}Montreal, Canada \\
\hspace{1.5cm}suryadipta.majumdar@concordia.ca}
\and
\IEEEauthorblockN{5\textsuperscript{th} Md. Shohrab Hossain}
\IEEEauthorblockA{\scriptsize
\textit{Department of CSE} \\
\textit{Bangladesh University of Engineering and Technology} \\
Dhaka, Bangladesh \\
mshohrabhossain@cse.buet.ac.bd}

\thanks{*Equal contribution.}
}

\maketitle

\begin{abstract}

With the rapid growth of smart home IoT devices, users are
increasingly exposed to various security risks, as evident from recent studies. 
While seeking answers to know more on those security concerns, users are mostly left with their own discretion while going through various sources, such as online blogs, and technical manuals; which may render  
higher complexity to the regular users to extract the necessary information from. 
This requirement 
does not go along with the common mindsets of smart home users and hence threatens the security of smart homes furthermore.
In this paper, we aim to identify and address the major user-level security concerns in smart homes.
Specifically,
we develop a novel dataset of Q\&A from public forums, capturing practical security challenges faced by smart home users. 
We extract major security concerns in smart homes from our dataset by leveraging the Latent Dirichlet Allocation (LDA).
We fine-tune relatively ``smaller'' transformer models, such as T5
and Flan-T5, on this dataset to build a QA system tailored for smart
home security. Unlike larger models like GPT and Gemini, which are powerful but often resource hungry, 
and requiring data sharing,
smaller models are more feasible for deployment in resource-constrained or privacy-sensitive environments, like smart homes. The dataset is manually curated and supplemented with
synthetic data to explore its potential impact on model performance.
This approach significantly improves the system’s ability to deliver accurate and relevant answers, helping users address common security concerns with smart home IoT devices. Our experiments on real-world user concerns show that our work improves the performance of the base models.

\end{abstract}

% \begin{IEEEkeywords}
% Smart Home Security, Transformer Models, Domain Specific Dataset
% \end{IEEEkeywords}

\section{Introduction}
The rapid adoption of smart home IoT devices has raised significant security and privacy concerns, including threats like intrusions and data theft \cite{vardakis2024review, sahu2024exploring}. Although manuals and online resources exist, their technical complexity often prevents users from finding clear guidance on securing their devices \cite{vetrivel2023examining}. This confusion can lead to critical oversights—for example, failing to change default settings on a smart camera—leaving homes exposed to unauthorized access.

With the rapid advancement of large language models (LLMs), we observe a significant trend in their adoption for various security applications (e.g., offensive security \cite{deng2024pentestgpt}, defensive security \cite{fang2024large}, generic cyber knowledge \cite{liu2024cyberbench}, generic IoT security \cite{chatiot}, artificial IoT \cite{aiot}). However, there is no study based on adopting LLMs especially in resource-constrained environments in smart home IoT security. Furthermore, there is a lack of datasets tailored to smart home IoT security that address user concerns \cite{zeng2017end}, making it difficult to evaluate the effectiveness of LLMs in such contexts. This gap highlights the need for research in assessing LLMs for smart home IoT security in resource-constrained settings. However, such study might encounter several challenges as follows:

% Existing literature on smart home IoT devices focuses mainly on identifying security and privacy concerns but lacks emphasis on providing practical guidance or solutions to help users address these issues, such as answering their specific questions and aiding them in securing their devices \cite{protick2024unveiling, vetrivel2023examining, Hasiuk2023SECURITYPI, emami2020informing}. Question-answering systems \cite{rajapaksha2024rag, krishna2024attackqa}, where the answer is generated from a provided context—often referred to as open-book question-answering systems—have shown success in fields like law \cite{zhong2020jec} and medicine \cite{lamurias2020generating}, but they cannot be directly applied to smart home security because they lack domain-specific training. General-purpose models and those trained in other areas struggle to address the specialized concerns of IoT environments, particularly in smart home security. Smaller models like T5, while efficient, are limited in providing accurate, domain-specific answers. Additionally, currently there is no data set available for fine-tuning systems specifically for smart home security concerns.

% These gaps highlight the need for our study, which fine-tunes transformer models using a novel, domain-specific dataset to provide precise answers tailored to the unique challenges of smart home environments.

% Despite the usefulness of such a tool to automatically answer the questions related to common concerns on smart home security, adopting the existing transformer-based popular models faces the following challenges:
\begin{itemize}[leftmargin=*, itemsep=0pt, topsep=0pt]
    \item[\textbullet] Building a smart home security dataset poses challenges due to the scarcity of public data \cite{touqeer2021smart} and the noisy, unstructured nature of forum content. Converting this raw data into clean, structured Q\&A pairs demands extensive preprocessing, noise filtering, and manual validation.

    % , which are resource-intensive processes.
   \item[\textbullet] Adapting generative models \cite{chung2024scaling} to smart home security is challenging, as they are pre-trained on broad QA datasets. Without domain-specific fine-tuning, they often yield incorrect or incomplete answers \cite{hallucinate1}, lacking the specialized knowledge needed for this nuanced domain. The scarcity of tailored public datasets worsens the problem.

    % , as such datasets are essential for effective fine-tuning and improving model performance.
\end{itemize}

Given these challenges, in this paper, we aim at building a framework for a question-answering system tailored to smart home security concerns. Specifically, we develop a structured dataset derived from real-world smart home security concerns, addressing the limitation of having a specific smart home IoT security-related dataset. We fine-tune different LLMs, specifically resource-efficient LLMs such as variants of T5 and Flan T5 \cite{chung2024scaling}, to enhance their capability of effectively answering domain-specific questions to help the users with resource-constraint environments and privacy concerns. Finally, we demonstrate the effectiveness of our framework compared to the base models in answering questions related to smart home security. 
% To overcome those challenges, we propose a framework for building a question-answering system tailored to smart home security concerns. We began by automatically collecting question-answer pairs from a diverse range of 19 public forums and communities to capture real-world user issues, employing Latent Dirichlet Allocation (LDA) \cite{blei2003latent} topic modeling  to identify and categorize prevalent topics. Next, we applied a semi-automatic approach, leveraging a large language model (LLM) to refine and enhance the quality of the dataset. Additionally, we explored a novel LLM-based technique to generate synthetic questions, constructing a synthetic question-answering dataset to supplement the original data. Finally, we fine-tuned multiple variants of T5 and Flan-T5 models, resulting in models that effectively answer domain-specific questions from context.
Our key contributions are:
% as follows:
\begin{itemize}[leftmargin=*, itemsep=0pt, topsep=0pt]
     \item 
     % \textbf{Creation and Refinement of a Dataset for Smart Home Security:}
     We develop a dataset of 3,319 question-answer pairs from a diverse range of 18 major public forums and communities along with several less popular sources (including platforms managed by popular providers, e.g.,  Google Nest \cite{google_nest_community}, Apple Community \cite{apple_community_discussions}, and Verizon \cite{verizon_community_forums}),
     focusing on real-world security concerns in smart homes. 
     % Specifically, we create different versions of dataset (i) Version 1.0 with raw question-answer pairs, (ii) Version 2.0 with more structured relevant question-answer pairs from Version 1.0, removing ambiguity and overly lengthy texts, and (iii) Version 3.0 with more compatible to fine-tuning question-answer pairs for resource-constrained models with relatively shorter answers while preserving the original meaning. 
     % Moreover, to create more variety of QA pairs and supplement the original data, we create a synthetic dataset using an LLM-based approach using Gemini-1.5-Flash \cite{Google_AI} based on original question-answer pairs. This dataset offers insights such as user perspectives, including common device misconfigurations and frequently faced security vulnerabilities, helping users and researchers to develop targeted solutions. 
     This dataset is publicly available at the following link\footnote{\url{https://github.com/IoTSmart-art/smarthome}} to support further research in this area.

    % \item \textbf{Creation and refinement of a specialized dataset:} We developed a dataset of 3,321 question-answer pairs from a diverse range of 19 public forums and communities, focusing on real-world smart home IoT security concerns. This dataset offers insights such as user perspectives, including common device misconfigurations and frequently faced security vulnerabilities, helping researchers develop targeted solutions. It will be made publicly available upon publication to support further research in this area.
    \item 
    % \textbf{Identification of Common Security Concerns in Smart Homes:} 
    To comprehensively identify prevalent security concerns in smart homes, we employ Latent Dirichlet Allocation (LDA) \cite{blei2003latent} to identify common topics from question-answer pairs gathered from multiple sources in our dataset. This analysis informs the enhancement of our question-answering system by targeting key issues while also offering valuable insights for researchers and developers to improve security solutions and product designs.
    % \item \textbf{Identification of common concerns using LDA topic modeling:} We employed LDA \cite{blei2003latent} to identify common security concerns in smart home IoT environments from question-answer pairs. This analysis informs the enhancement of our question-answering system by targeting key issues, while also offering valuable insights for researchers and developers to improve security solutions and product design.
    \item 
    % \textbf{Fine-tuning ``Smaller'' LLMs for Smart Home Security:} 
    We fine-tune different variants of T5 (T5-small, T5-base) and Flan T5 models (Flan-T5-small, Flan-T5-base, Flan-T5-large) due to their low resource requirements and potential suitability in the smart home context. This fine-tuning, using our dataset.
    % , aim at enhancing their ability to answer questions related to security concerns in smart homes. 
    This adaptation addresses the challenge of the absence of QA systems tailored to this complex and evolving field, enhancing users' ability to access practical solutions to their security concerns.
    % \item \textbf{Fine-tuning transformer models for smart home security:} By fine-tuning multiple transformer models on our novel dataset specific to smart home IoT device security and privacy, we enabled effective domain-specific question answering. This adaptation addresses the challenge of the absence of QA systems tailored to this complex and evolving field, enhancing users' ability to access practical solutions to their security concerns.
  \item 
  % \textbf{Demonstration of Framework Effectiveness:}
  Our experiments on real-world user concerns demonstrate the effectiveness of our framework, achieving the best overall results among all five models. 
  Additionally, we compare those fine-tuned models with several major base models (GPT-4o, GPT-4omini, Llama-3.3-70B, Llama-3.2-3B, Llama-3-70B, and Llama-3.1-70B).
  The F1 Score of our approach improves by 89.77\% (vs. 39.57\% in BERTScore, which measures semantic similarity between the generated and reference answers) in the validation set and 57.36\% (vs. 45.19\% in BERTScore) on the test set.

\end{itemize}

% The rest of the paper is organized as follows. In Section 2, we review the related works and outline their limitations. Section 3 presents the methodology of dataset creation and the fine-tuning process. Section 4 discusses the experimental results and findings. Section 5 reports our key observations and their potential solutions. Section 6 concludes the paper with a discussion of future work. 

\section{Related Works}

This section reviews the related work.

% \subhead{Studying Security Concerns in Smart Homes}
As smart home technologies expand, so do concerns around security and privacy \cite{guhr2020privacy}. Interconnected devices are vulnerable to breaches, and users often prioritize convenience over security. Studies show that many struggle to find reliable guidance \cite{vetrivel2023examining}, while technical issues and unclear documentation further complicate secure usage. As a result, users rely on ad hoc strategies and lack practical resources. To address this gap, our work builds a curated dataset and fine-tuned QA system to offer actionable solutions.

% \subhead{Development of Generative QA Models in Other Domains}
Generative QA systems have made notable progress in domains like law \cite{zhong2020jec} and medicine \cite{yadav2022chq}, where domain-specific knowledge is crucial. Datasets like CHQ-SUMM \cite{yadav2022chq} show how well-curated, specialized data can improve QA accuracy. In cybersecurity, prior work includes QA over generic domain knowledge \cite{tihanyi2024cybermetric}, classification tasks \cite{liu2024cyberbench}, and RAG-based systems for IoT security \cite{chatiot}. Unlike them, in this work, we focus on developing a generative question-answering specialized for smart home security, mainly from the user perspective. 

\begin{table*}[ht] 
\scriptsize
\caption{Comparison of existing works with ours. The symbols (\circleTick), (\circleNon) and (-) mean supported, not supported and not applicable, respectively.}
\centering 
\renewcommand{\arraystretch}{0.8} % Reduce row height
\setlength{\tabcolsep}{3pt} % Reduce column separation
% \resizebox{\textwidth}{!}{%
\begin{tabular}{|l|c|c|c|c|c|c|c|c|c|c|c|c|c|c|} 
\hline 
\multirow{2}{*}{\textbf{Criteria}} & 
\multicolumn{4}{c|}{\textbf{Application}}  & 
\multicolumn{2}{c|}{\textbf{Type}} & 
\multicolumn{3}{c|}{\textbf{Dataset Creation}} & 
\multicolumn{5}{c|}{\textbf{Evaluation Strategy}} \\ \cline{2-15}
&
\rotatebox{90}{\parbox{2cm}{\scriptsize Smart Home\\IoT Security}} & 
\rotatebox{90}{\parbox{1.5cm}{\scriptsize Artificial IoT}} & 
\rotatebox{90}{\parbox{2cm}{\scriptsize Generic IoT\\Security}} & 
\rotatebox{90}{\parbox{1cm}{\scriptsize Others}} & 
\rotatebox{90}{\parbox{1.8cm}{\scriptsize Non Open\\Ended}} & 
\rotatebox{90}{\parbox{1.5cm}{\scriptsize Open Ended}} & 
\rotatebox{90}{\parbox{2.2cm}{\scriptsize Multi Source\\Data Collection}} & 
\rotatebox{90}{\parbox{1.5cm}{\scriptsize Data\\Refinement}} & 
\rotatebox{90}{\parbox{2cm}{\scriptsize Synthetic Data\\Generation}} & 
\rotatebox{90}{\parbox{1.8cm}{\scriptsize Base LLM\\Evaluation}} & 
\rotatebox{90}{\parbox{2cm}{\scriptsize Evaluation with\\Finetuning}} & 
\rotatebox{90}{\parbox{2.2cm}{\scriptsize Effect of Dataset\\Refinement}} & 
\rotatebox{90}{\parbox{2cm}{\scriptsize Effect of\\Synthetic Data}} &
\rotatebox{90}{\parbox{1.5cm}{\scriptsize Effect of\\RAG}} \\[1ex]
\hline
AIoT \cite{aiot} &
\circleNon & \circleTick & \circleNon & \circleNon & \circleTick & \circleNon & \circleTick & \circleNon & \circleNon & \circleTick & \circleTick & \circleNon & \circleNon & \circleNon \\
\hline
Cyberbench \cite{liu2024cyberbench} &
\circleNon & \circleNon & \circleNon & \circleTick & \circleTick & \circleNon & \circleTick & \circleNon & \circleNon & \circleTick & \circleTick & \circleNon & \circleNon & \circleNon \\
\hline
SecureBERT \cite{aghaei2023securebert} &
\circleNon & \circleNon & \circleNon & \circleTick & \circleTick & \circleNon & \circleTick & \circleNon & \circleNon & \circleTick & \circleTick & \circleNon & \circleNon & \circleNon\\
\hline
Secqa \cite{liu2023secqa} &
\circleNon & \circleNon & \circleNon & \circleTick & \circleTick & \circleNon & \circleNon & \circleNon & \circleNon & \circleTick & \circleNon & \circleNon & \circleNon & \circleNon \\
\hline
WMDP \cite{wmdp} &
\circleNon & \circleNon & \circleNon & \circleTick & \circleTick & \circleNon & \circleTick & \circleNon & \circleNon & \circleTick & \circleTick & \circleNon & \circleNon & \circleNon \\
\hline
Cybermetric \cite{tihanyi2024cybermetric} &
\circleNon & \circleNon & \circleTick & \circleTick & \circleTick & \circleNon & \circleTick & \circleNon & \circleNon & \circleTick & \circleNon & \circleNon & \circleNon & \circleNon \\
\hline
ChatIoT \cite{chatiot} &
\circleNon & \circleNon & \circleTick & \circleNon & \circleNon & \circleTick & - & - & - & \circleTick & \circleNon & \circleNon & \circleNon & \circleTick \\
\hline
\textbf{Ours} &
\circleTick & \circleNon & \circleNon & \circleNon & \circleTick & \circleNon & \circleTick & \circleTick & \circleTick & \circleTick & \circleTick & \circleTick & \circleTick & \circleNon \\
\hline
\end{tabular}
% }
\label{tab:comparison}
\end{table*}

% \subhead{Comparison with Other Works} 
Table \ref{tab:comparison} highlights key differences between existing works and ours across applications, question types, dataset creation, and evaluation strategies. Open-ended questions, as used in models like ChatIoT \cite{chatiot}, lack specific ground-truth answers and are evaluated subjectively. In contrast, our non-open-ended questions have well-defined answers, enabling systematic evaluation based on accuracy and relevance. Unlike existing works, we first develop a dataset focused on smart home IoT security, then fine-tune resource-constrained LLMs to build a generative QA model. We evaluate model performance across different dataset versions (with and without refinement) and further examine the impact of synthetic data generation.

\section{Methodology}
This section presents our methodology.
% and then elaborates on each step of it.

% https://docs.google.com/presentation/d/1rKHJicjRMyfTxYIBJ7oUDOaLl261Kxb-RPlOJPYMCnM/edit?usp=sharing

% https://docs.google.com/presentation/d/1gyuiJ0oijC_T3_Wo0tNH2GLvpllk4TxuAuXhfQRINIM/edit?usp=sharing

\subsection{Overview} 
\begin{figure*}[ht]
    \centering
    \includegraphics[width=\textwidth]{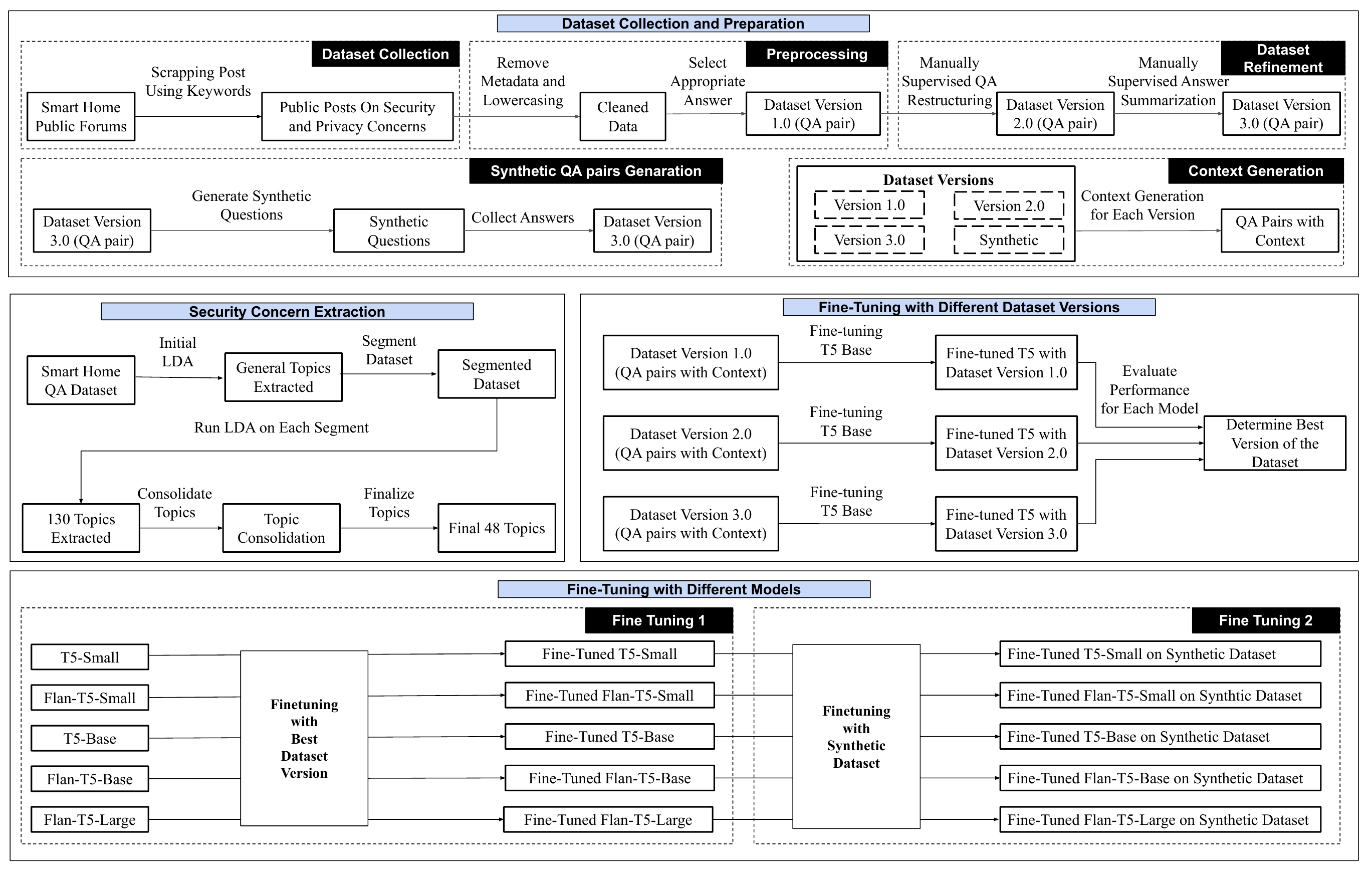}
    \caption{An overview of our proposed approach.}
    \label{fig:approach_overview}
\end{figure*}

Figure \ref{fig:approach_overview} illustrates an overview of our approach. 
First, we collect posts related to smart home security and privacy concerns from various public forums, using targeted keywords to identify relevant content. 
Second, we preprocess to convert these posts into question-answer (QA) pairs to create the initial dataset (Version 1.0). Third, we refine this original dataset into two more refined versions (Version 2.0 and Version 3.0). 
Fourth, to facilitate fine-tuning, we generate context for each question-answer pair through a semi-automated process involving a large language model (LLM). 
Fifth, we use the LLM to create a synthetic question dataset derived from the original dataset and then collect corresponding answers to construct a synthetic question-answer dataset. 
Sixth, we conduct topic analysis using LDA to identify the primary security topics within the dataset. 
Seventh, we evaluate the effectiveness of our dataset refinement strategy by fine-tuning a T5-base model using all three versions of the dataset. Finally, we fine-tune various T5 and Flan-T5 models to enable them to answer questions from the context. 
% We evaluate their performance across multiple test sets to determine the improvements and effectiveness of our models. 
% \begin{figure}[ht]
%     \centering
%     % \includegraphics[width=0.95\textwidth]{Images/Flowchart.png}
%     \includegraphics[width=\textwidth]{NEW_One/overview.pdf}
%     \caption{An overview of our proposed approach.}
%     \label{fig:approach_overview}
% \end{figure}

% Our primary goal is to develop a generative question-answering system using transformer models that can effectively address questions within the domain of smart home security and privacy. We began by collecting question-answer pairs from public forums and preprocessing them to create the original dataset, Version 1.0. We generated the necessary context for each question-answer pair using a large language model (LLM) and then refined the dataset to produce two improved versions. We specifically assessed how the quality of these refined datasets influenced model performance. Additionally, we created a synthetic dataset using the LLM to evaluate whether it could further enhance the model's effectiveness. Finally, we fine-tuned different variants of T5 and FLAN-T5 models and tested their maximum performance on both validation and test sets, using metrics such as F1-score, BERTScore F1, and ROUGE-L to measure the results, thus producing a domain-specific transformer model capable of answering questions accurately.

% For evaluation, we relied on metrics such as F1-score, BERTScore F1, and ROUGE-L.

% \subsection{Converting Commonly Raised Security Concerns in Smart Homes
% to a Corpus of Question-Answer (QA) Pairs}

\subsection{Data Collection and Preparation}
% In the following, 
We describe how we collect and prepare the data as follows.

% \subsubsection{Extracting Common Smart Home Security Concerns from Public Forums}
\subhead{Data Collection} \label{sec:dataset_collection}
To develop a comprehensive dataset of QA pairs related to smart home IoT security, it is crucial to analyze real-world discussions where users actively share their experiences, ask questions, and seek solutions. Public forums and communities serve as rich sources of user-generated content, offering diverse insights into real-world problems, concerns, and knowledge gaps. By analyzing these discussions, we can identify trends, recurring issues, and user expectations related to smart home security and privacy. 
We collect QA pairs from a diverse range of 18 major public forums and communities, including platforms like AVS \cite{avs_forum}, DIY Home \cite{diychatroom_forum}, CocoonTech \cite{cocoontech_forum}, Digital Home \cite{digitalhome_forum}, DIYNot \cite{diynot_forum}, Ezlo \cite{ezlo_forum}, Home Assistant \cite{home_assistant_forum}, Reddit \cite{reddit_forum}, SmartThings \cite{smartthings_forum}, SNB 
 \cite{snb_forum}, Stack Exchange \cite{stackexchange_iot}, OpenWRT \cite{openwrt_forum}, level1techs \cite{level1techs_forum}, and Tom’s Guide \cite{feedspot_automation_forum}, as well as communities associated with companies like Google Nest \cite{google_nest_community}, Apple Community \cite{apple_community_discussions}, and Verizon \cite{verizon_community_forums} and whirlpool \cite{whirlpool_forum} and other less popular sources such as \cite{quora_forum}. 
We choose these public forums because they are popular platforms where users seek solutions to their problems related to smart home devices. To find posts relevant to smart home security and privacy concerns, we derive a targeted list of keywords. Using this curated list, we scrape posts from these forums, adhering to their guidelines. This method ensures that we collect relevant content and capture a broad range of discussions, contributing to a comprehensive dataset on smart home security issues. The distribution of QA pairs collected from different sources is summarized in Table \ref{tab:qa_statistics}.

\begin{table*}[t]
    \caption{Sources of QA pairs used in our dataset.}
    \centering
    \scriptsize
    \renewcommand{\arraystretch}{1.3}
    \setlength{\tabcolsep}{2pt}
    \resizebox{\textwidth}{!}{%
    \begin{tabular}{|l|*{19}{c|}}
        \hline
        \makecell{\textbf{Differnt}\\\textbf{Forums}\\} & 
        \makecell{AVS\\Forum\\\cite{avs_forum}} & 
        \makecell{Smart\\Things\\\cite{smartthings_forum}} & 
        \makecell{Home\\Assistant\\\cite{home_assistant_forum}} & 
        \makecell{Ezlo\\\cite{ezlo_forum}} & 
        \makecell{Cocoon\\Tech\\\cite{cocoontech_forum}} & 
        \makecell{Other\\Forums\\\cite{quora_forum}} & 
        \makecell{Digital\\Home\\\cite{digitalhome_forum}} & 
        \makecell{DIY\\Not\\\cite{diynot_forum}} & 
        \makecell{Whirl\\pool\\\cite{whirlpool_forum}} & 
        \makecell{Google\\Nest\\\cite{google_nest_community}} & 
        \makecell{Apple\\Community\\\cite{apple_community_discussions}} & 
        \makecell{Verizon\\\cite{verizon_community_forums}} & 
        \makecell{level1\\techs\\\cite{level1techs_forum}} & 
        \makecell{Open\\WRT\\\cite{openwrt_forum}} & 
        \makecell{DIY\\Home\\\cite{diychatroom_forum}} & 
        \makecell{Red\\dit\\\cite{reddit_forum}} & 
        \makecell{SNB\\\cite{snb_forum}} & 
        \makecell{Tom’s\\Guide\\\cite{feedspot_automation_forum}} & 
        \makecell{Stack\\Exchange\\\cite{stackexchange_iot}} \\
        \hline
        QA Collected & 6,922 & 6,576 & 6,240 & 4,353 & 1,882 & 1,700 & 1,482 & 1,446 & 1,430 & 1,320 & 1,220 & 1,150 & 1,080 & 900 & 734 & 285 & 229 & 48 & 46 \\
        \hline
        QA Selected  & 218   & 500   & 481   & 356   & 185   & 150   & 64    & 164   & 120   & 110   & 100   & 145   & 140   & 74  & 96  & 285 & 67  & 38  & 26 \\
        \hline
    \end{tabular}%
    }
    \label{tab:qa_statistics}
\end{table*}

% \subsubsection{Preparing the question-answer pairs from security concerns}
\subhead{Preprocessing}
The raw data collected from these forums and communities are often unstructured. They include different types of unnecessary data, including different types of meta data (e.g., votes, dates, etc.) that are not necessary for preparing the dataset. To get a cleaner set of QA pairs, the collected  data from the public forums undergoes several preprocessing steps to ensure a clean, structured dataset.

\noindent \textit{Step 1 - Cleaning and Formatting the Data: } After scraping using no-code web scraping tools like ParseHub \cite{parsehub} and Octoparse \cite{octoparse}, our main goal is to build a dataset containing only QA pairs. The scraping process includes a significant amount of metadata, such as dates, times, authors, and votes, which are not needed for our dataset. Therefore, we discard this metadata to focus solely on the question-answer content. Additionally, we standardize all text to lowercase to prevent case sensitivity from adding complexity during the fine-tuning stage, ensuring consistency and simplifying downstream processing tasks like text matching and analysis.

\noindent \textit{Step 2 - Answer Selection: } After isolating the question-answer pairs, we find that many questions have multiple answers. For our dataset, we require only one answer per question. To select the most appropriate answer, our criterion is length—we choose the longest answer to ensure a comprehensive response, enhancing dataset quality and reducing redundancy.

While this method may preserve longer, less precise answers, Versions 2.0 and 3.0 apply Gemini-1.5-Flash with manual review to enhance clarity and correctness. Inaccuracies and outdated content are revised to ensure a high-quality, reliable dataset for fine-tuning.

\begin{figure}[ht]
    \centering
    \includegraphics[width=0.5\textwidth]{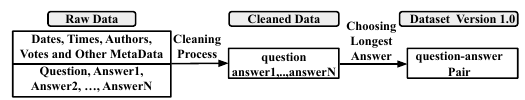}
    \caption{An example of the preprocessing step of dataset creation.}
    \label{fig:preprocessing}
    \vspace{-18pt}
\end{figure}

\noindent \textit{Example 1:} Figure \ref{fig:preprocessing} illustrates an example of our preprocessing step. On the leftmost side, we have our collected raw data, including multiple non-essential metadata (such as dates, times, authors, votes, etc.), with one question and multiple answers ($answer_1$, ..., $answer_N$). In the next step, we remove all the meta data and lowercase the questions and answers to have a more cleaned version of the dataset with only the questions and the answers. After that, we choose the longest answer from the available answers to have a comprehensive response for the question. 
Finally, we consolidate the refined data, reducing the original 39,043 question-answer pairs to 3,319 high-quality pairs to create Version 1.0 of the dataset.

% This reduction, through answer selection, focuses on enhancing the integrity and relevance of the data set, ensuring that it is both manageable and perfectly suited for further analysis and model training in the context of smart home security research.

\subhead{Dataset Refinement} \label{sec: dataset_refinement}
After creating the original dataset (Version 1.0), we develop two incrementally improved versions, Version 2.0 and Version 3.0, by leveraging the capabilities of Gemini-1.5-Flash \cite{Google_AI} alongside manual supervision. This approach significantly enhanced the dataset's quality through structured refinement and validation.

\noindent \textit{Rephrasing for Clarity and Consistency: }
% \paragraph{Creation of Dataset Version 2.0}
The original dataset (which we refer to as Version 1.0) has several issues. Since the QA pairs are collected from public forums, the questions are often unstructured, ambiguous, and difficult to understand. The answers tend to be overly lengthy, unstructured, and sometimes unclear. For effective fine-tuning of transformer models, we need clear, concise, and well-structured question-answer pairs. To improve clarity, we use Gemini-1.5-Flash to reformulate the questions and restructure the answers. After the automated process, we manually review the dataset to correct inconsistencies. Thus, from Version 1.0, we create Version 2.0 of the dataset, which is more structured, containing concise and easy-to-understand question-answer pairs.

Manual review was feasible for all 3,319 QA pairs in Versions 2.0 and 3.0. Gemini-1.5-Flash provided structured drafts, allowing reviewers to efficiently verify accuracy and clarity.

\begin{figure}[ht]
    \centering
    \includegraphics[width=0.5\textwidth]{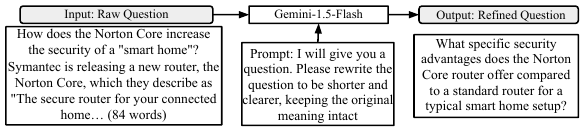}
    \caption{An example of the restructuring of questions.}
    % using Gemini-1.5-Flash \cite{Google_AI} to create Version 2.0 of the dataset from Version 1.0.}
    \label{fig:question_restructure}
    \vspace{-18pt}
\end{figure}

\noindent \textit{Example 2.} Figure \ref{fig:question_restructure} provides an example of this refinement process. On the left side, we present a raw question (44 words long) from Version 1.0 of our dataset. We give it as input to Gemini-1.5-Flash with a specific prompt to refine the question. On the right side, we obtain a refined question (22 words long), reducing its length while improving clarity and conciseness. Furthermore, a manual review ensures that, despite reformulation and restructuring, the original meaning of the questions and answers is preserved.

\noindent \textit{Summarizing Long Answers: }
In Version 2.0, we still encounter the issue of lengthy answers. Although the length is reduced compared to Version 1.0, it remains relatively long. Due to computational constraints and our use of smaller LLMs, we decide to summarize the answers, resulting in a more compact dataset. Thus, from Version 2.0, we create Version 3.0 of the dataset by using Gemini-1.5-Flash to condense the answers while preserving their core meaning. 

\begin{figure}[ht]
    \centering
    \includegraphics[width=0.5\textwidth]{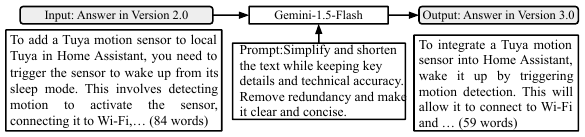}
    \caption{An example of the process of summarizing the answer.}
    % using Gemini-1.5-Flash \cite{Google_AI}.}
    \label{fig:summarize}
    \vspace{-10pt}
\end{figure}

\noindent \textit{Example 3.} Figure \ref{fig:summarize} shows an example of the summarizing process. The left side contains a lengthy answer from Version 2.0, which serves as the original input to the Gemini-1.5-Flash model. The model is given a specially crafted structured prompt to generate a shorter version of the original input, as indicated. The intention behind this question is to shorten the length of the answer, while simultaneously making sure that the shortened output is still able to retain the original purpose and meaning expressed in the longer version. The right side shows a shorter version of the answer compared to the input. 

Table \ref{tab:length} provides a summary of the average QA lengths in different versions of the dataset, showing the progression to greater clarity and conciseness of the QA pairs.

\begin{table}[ht]
\caption{Avg. question and answer lengths (in words).}
\scriptsize
\centering
\begin{tabular}{|c|c|c|}
\hline
\textbf{Version} & \textbf{Avg. Question Length} & \textbf{Avg. Answer Length} \\ \hline
1.0 & 149.80 & 462.47 \\ \hline
2.0 & 20.00 & 159.70 \\ \hline
3.0 & 20.00 & 40.45 \\ \hline
\end{tabular}
\label{tab:length}
\end{table}
\subhead{Synthetic QA Pair Generation}
To supplement the original dataset and increase QA variation, we generate a synthetic dataset using large language models (LLMs). Our goal is to assess whether LLMs alone can create high-quality new questions that enhance dataset diversity. While traditional methods include rule-based transformations \cite{keklik2019rule}, template-based generation \cite{fabbri2020template}, and paraphrasing, we adopt a semi-automated approach using an LLM to efficiently generate diverse QA pairs from existing data.

We prompt the LLM to generate new, contextually distinct questions based on the core ideas of original QA pairs. These go beyond paraphrasing to explore new angles. Answers are then sourced from trusted online materials such as forums, technical documentation, and manuals, resulting in 2,245 synthetic QA pairs.

\begin{figure}[ht]
    \centering
        \vspace{-10pt}
    \includegraphics[width=0.5\textwidth]{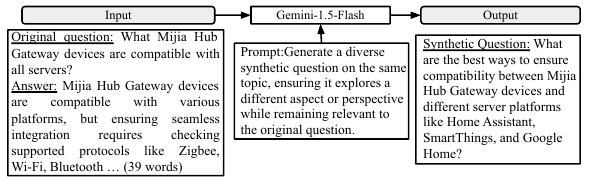}
        \vspace{-15pt}
    \caption{An example of the generation of synthetic questions.}
    \label{fig:synthetic}
    \vspace{-5pt}
\end{figure}

\textit{Example 4.} Figure \ref{fig:synthetic} shows an example of how to generate synthetic questions using Gemini-1.5-Flash. An original QA pair is provided as input, and the model, guided by a specific prompt, generates a synthetic question that explores a different aspect of the original topic. This process helps introduce meaningful variation and enhance the dataset's richness.

\subhead{Context Generation}
To complete our dataset, we generate context for each QA pair. In generative QA, context provides the background needed for accurate and informed answers \cite{kabongo2024effective}, and is essential for fine-tuning. We adopt a semi-automated approach using Gemini-1.5-Flash, which generates initial context for each QA pair. These outputs are then manually reviewed and refined using reliable sources such as blogs and technical manuals to ensure accuracy.

\begin{figure}[ht]
    \centering
    \includegraphics[width=0.5\textwidth]{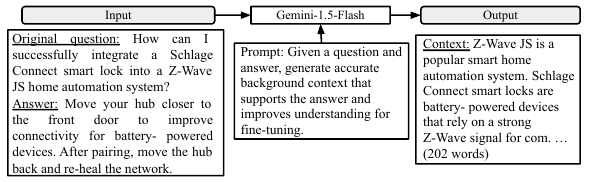}
    \caption{An example of illustrating the context generation}
    % process for a QA pair using Gemini-1.5-Flash \cite{Google_AI}.}
    \label{fig:context}
    \vspace{-15pt}
\end{figure}

\noindent \textit{Example 5.} Figure \ref{fig:context} shows an example where the original question \textit{``How can I successfully integrate a Schlage
Connect smart lock into a Z-Wave JS home
automation system?''} as well as the corresponding answer, \textit{``Move your hub closer to the front door to
improve connectivity for battery-powered
devices. After pairing, move the hub back and
re-heal the network.''} of the question as input to the Gemini-1.5-Flash with a specific prompt to generate the corresponding context. The Gemini-1.5-Flash generates a context based on the question and answer it gets as input. We manually review the generated context to ensure its quality and relevance, thus making it prepared to finetune the models with it.

% \subsection{Dataset Analysis}
\subsection{Security Concern Extraction}

\subhead{Initial LDA Modeling and Dataset Overview}
To identify prevalent concerns in our smart home IoT security dataset, we apply Latent Dirichlet Allocation (LDA) \cite{blei2003latent} for topic modeling. LDA is an unsupervised probabilistic method that models each document as a mixture of latent topics, with each topic represented by a word distribution. It uses a Bayesian framework with Dirichlet priors to uncover hidden topic structures through term co-occurrence patterns. Well-suited for large, unstructured datasets like public forums, LDA reveals recurring themes without the need for labeled data.

This analysis serves two main purposes: to refine and target the question-answering system to address key issues more effectively, and to provide valuable insights for researchers and developers looking to enhance security solutions and improve product design.

Our initial run applied LDA across the entire dataset to extract dominant themes. While this yielded broad categories such as “security” and “privacy,” the results lacked granularity and often merged distinct concerns. This limitation motivated a more segmented and targeted analysis, detailed in the next section.

\subhead{Detailed LDA Analysis and Topic Summarization}
To gain deeper insights into smart home security concerns, we refine our initial LDA analysis by segmenting the dataset into 13 parts. As the data spans 18 forums with distinct discussion patterns, we apply LDA separately to each segment. The first 12 subsets correspond to forums with the highest number of questions, capturing the most active discussions. The 13th subset combines data from smaller communities, such as SNB \cite{snb_forum}, to capture similar but less frequent security-related topics.

Key extracted themes include Device and Network Security: addressing unauthorized access, encryption concerns, and firmware vulnerabilities in IoT devices; Home Security Systems and Camera Integration: exploring security challenges in surveillance systems, access control, and privacy concerns related to cloud-based storage; Smart Home System Customization: highlighting user preferences for security configurations, automation safety, and integration complexities with multiple IoT platforms; and IoT Device Security and Network Vulnerabilities: covering common attack vectors such as botnets, weak authentication mechanisms, and risks associated with insecure communication protocols \cite{vardakis2024review}. Some of the final topics identified through LDA analysis, along with their associated keywords, are summarized in Table \ref{tab:lda_topics}.

\begin{table}[ht]
\caption{Final topics identified with keywords.}
\scriptsize
\centering
\begin{tabular}{|p{3cm}|p{5cm}|}
\hline
\textbf{Topic} & \textbf{Keywords} \\
\hline
Device and Network Security & device, network, router, security, smart, wifi, vpn, VLAN, privacy, access, attack, hack, malware \\
\hline
Home Automation Systems and Control & home, system, control, automation, smart, device, switch, light, tv, video, usage, integration, monitoring, app \\
\hline
Network Configuration and Optimization & router, network, configuration, DNS, setup, google, wifi, IP, advanced settings, speed, performance, NUC, mesh, connection \\
\hline
Smart Home Device Integration and Management & smart, device, home, integration, connect, google, speaker, service, system, expand, configuration, app \\
\hline
Home Security Systems and Camera Integration & home, security, camera, system, control, smart, sensor, alarm, door, window, monitoring, app, lock, ADT, SmartThings \\
\hline
Home Automation and Power Management & home, power, system, control, switch, thermostat, energy, meter, lighting, automation, need, want \\
\hline
Software and System Integration & software, system, integration, smart, home, microsoft, program, version, install, debug, load \\
\hline
Device Connectivity and USB Management & usb, control, tv, drive, file, system, mac, connect, access, storage, memory \\
\hline
Home Services and Provider Integration & service, home, system, area, certification, control, canadian, Roger, monitoring, broadcasting, satellite, Bell, ExpressVu \\
\hline
Smart Home System Customization and Preferences & customization, home, security, system, door, camera, window, smart, want, integration, cost \\
\hline
\end{tabular}

\label{tab:lda_topics}
\end{table}

\subsection{Fine-Tuning with Different Dataset Versions} \label{sec:finetune_with_different}

To evaluate the impact of dataset quality on model performance for question-answering tasks, we fine-tune the T5-base model using three different training sets derived from the three refined versions of the dataset (Version 1.0, Version 2.0, and Version 3.0). Each training set contains 2,383 question-answer pairs, but the level of refinement varied based on the dataset version. For each training set, the question-answer pairs are taken from the corresponding dataset version, with Version 1.0 containing the original pairs, Version 2.0 including their improved versions, and Version 3.0 featuring the most refined versions. This means that while all three training sets contained the same underlying question-answer pairs, they differ in terms of clarity, completeness, and structural refinement. This approach ensures that the model is trained on progressively enhanced versions of the dataset, allowing us to assess how improvements in data quality influenced model performance.

For evaluation, we use a testset of 340 QA pairs exclusively from Version 3.0. Additionally, a validation set of 596 QA pairs is used to fine-tune the model before final testing. In total, the dataset consists of 3,319 question-answer pairs, with 2,383 allocated for training, 596 for validation, and 340 for testing. This decision ensures that the final assessment is conducted on the most structured and refined dataset, providing a reliable benchmark for performance comparison. The dataset is tokenized by concatenating the question and context into a single input sequence, paired with the corresponding answer as the target output. The same hyperparameters are applied across all three training sets to maintain consistency, balancing computational cost and model performance.

\subsection{Fine-tuning with Different Models}
% \subsection{Fine-tuning Different Models for Optimal Performance}
This step is to evaluate the best achievable performance of different models using our refined dataset. Specifically, we utilize Version 3.0 of the dataset (detailed in Section \ref{sec:refining}) for this purpose. We divide Version 3.0 into 2,383 QA pairs for training, 596 for validation, and 340 for testing purpose.

% In Fine-tuning 1 step, as illustrated in Figure \ref{fig:approach_overview}, 
We fine-tune 5 variants in total, from both T5 and Flan-T5, following the same fine-tuning process discussed in the previous section \ref{sec:finetune_with_different}. In the first stage, all models are fine-tuned using Version 3.0. Subsequently, we perform a second round of fine-tuning using the synthetic dataset to assess any potential performance improvements. The synthetic dataset consists of 1,792 question-answer pairs for training and 453 for validation, totaling 2,245 QA pairs. For testing, we use the same test dataset from the earlier stage to maintain consistency in evaluation. The same hyperparameters are applied consistently across all models during this stage. Finally, after completing both stages of fine-tuning, we evaluate the models' performance using the test sets, calculating the relevant metrics to determine the effectiveness of each model variant.
After fine-tuning, the models are deployed on Hugging Face\cite{huggingface} and made accessible via inference APIs, enabling real-time integration into applications for smart home security queries. This allows developers and researchers to easily use the models for domain-specific tasks.

\begin{figure*}[ht]
    \centering
    \includegraphics[width=0.8\textwidth]{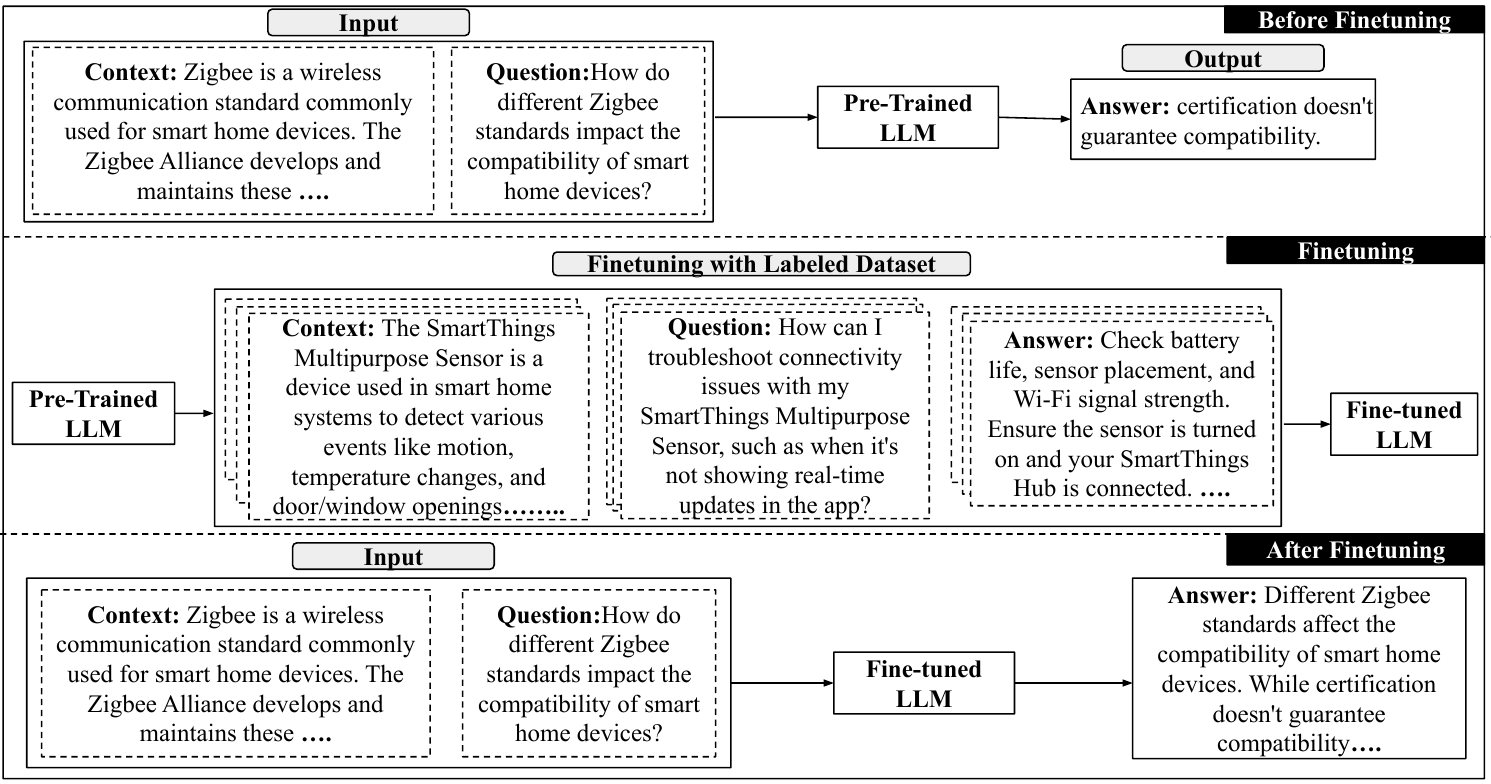}
    \vspace{-0.1in}
    \caption{Example of the finetuning LLM and performance development.}
    \vspace{-0.1in}
    \label{fig:finetuningLLM}
\end{figure*}

\noindent \textit{Example 6.} Figure \ref{fig:finetuningLLM} illustrates the fine-tuning process. In the first box, we provide context and our question \textit{``How do different Zigbee standards impact the compatibility of smart home devices''} as an input to the LLM. Before finetuning, the LLM cannot provide us with any meaningful answer, such as: \textit{``certification doesn't guarantee compatibility.''}. In the middle box, we finetune the pre-trained LLMs with our labeled dataset that includes the context, corresponding questions, and appropriate answers. This process results in a fine-tuned LLM. In the last box, we provide the same input as before. This time, the finetuned LLM generates a more meaningful answer: \textit{``Different Zigbee standards affect the compatibility of smart home devices. While certification doesn't guarantee compatibility….''} from the input provided.

\section{Experiments}

% \textcolor{red}{We evaluate the impact of dataset quality on model performance and experiment with various T5 and FLAN-T5 to measure their performance after finetuning. Additionally, we explore the effectiveness of our synthetic dataset. The results of these experiments are presented here.} In our study, we investigate three research questions (RQs): 

This section evaluates the effectiveness of this work.
% by answering the following research questions:
% \begin{itemize}[leftmargin=*, itemsep=0pt, topsep=0pt]
%     \item \textbf{RQ1: } What is the impact of dataset refinement on model performance?
%     \item \textbf{RQ2: } What is the impact of two-stage fine-tuning on T5 and Flan-T5 model variants?
%     \item \textbf{RQ3: } How does our fine-tuned model compare to other LLMs in addressing smart home security concerns?
% \end{itemize}

\subsection{Experimental Setup}
% We evaluate the impact of dataset quality on model performance and experiment with various T5 and FLAN-T5 variants using the best dataset version to determine the maximum achievable performance. Additionally, we explore the effectiveness of our synthetic dataset. The results of these experiments are presented here.

\subhead{Experimental Environment}
We train our models in a cloud environment with an \textit{NVIDIA T4 GPU}, using pretrained T5 variants from the Hugging Face Model Hub \cite{huggingface}. Fine-tuning is conducted with the Hugging Face Transformers \cite{huggingface} library via the Trainer API. We also use Langchain \cite{LangChain} to evaluate other LLMs.

\subhead{LLMs and Parameter Selection} In our experiments, 
% we evaluate different variants T5 and Flan-T5 models, including base and fine-tuned models. Moreover, to compare the performance between our fine-tuned models and other open source or closed source models, we evaluate GPT-4o and GPT-4omini through OpenAI API \cite{OpenAI}, different variants of Llama Models \cite{Llama} (Llama-3.2-3B, Llama-3-70B, Llama-3.1-70B, Llama-3.3-70B) via Fireworks API \cite{fireworks}. 
We set temperature and seed to 0 for deterministic outputs, following prior work \cite{alam2024ctibench}, and limit outputs to 512 tokens. T5 and Flan-T5 variants are fine-tuned with tuned hyperparameters to balance performance and efficiency.

% The key training parameters are as follows:  

% The learning rate is set to $3 \times 10^{-5}$, with a batch size of 4 for both training and evaluation. To prevent overfitting, a weight decay of 0.01 is applied. The training is conducted over four epochs, with early stopping enabled to halt training if no improvement is observed for one consecutive epoch. Gradient accumulation is set to 4 steps to efficiently manage memory constraints. Model evaluation is performed at the end of each epoch, and logging occurs every 50 steps to track training progress. Checkpoints are saved at the end of each epoch, with a retention limit of 3 to conserve storage while maintaining model progression. The best-performing model is selected based on validation loss (\texttt{eval\_loss}), and the system is configured to automatically load this optimal checkpoint for final deployment. To ensure reproducibility and accessibility, the fine-tuned models are integrated with the Hugging Face Model Hub \cite{huggingface} for deployment and further experimentation.  

\subhead{Evaluation Metrics}
We select three metrics to assess model performance at the lexical, structural, and semantic levels: 
F1-Score \cite{rajpurkar2016squad}, ROUGE-L \cite{lin2004rouge},
and BERTScore (F1) \cite{zhang2019bertscore}. F1 measures precision and recall, ROUGE-L captures structural similarity, and BERTScore ensures semantic alignment. Together, they provide a comprehensive evaluation of QA performance.

\subsection{Impact of Dataset Refinement on Model Performance} \label{sec:refining}

In this section, we evaluate the effectiveness of our semi-automated refining approach (detailed in Section \ref{sec: dataset_refinement}) in enhancing the quality of our dataset so that the model can be better fine-tuned for answering domain-specific questions from the context of smart home IoT security. To achieve this, we fine-tune a T5-base model using all three versions of the dataset individually and assess its performance on the test set. The evaluation is conducted using three key metrics: F1 Score, ROUGE-L Score, and BERT F1 Score, as described earlier. 

On the testset, the F1-score shows limited improvement after fine-tuning with Dataset Version 1.0 due to noise and unclear question-answer pairs, increasing only from 0.3500 (base Model) to 0.4280. With Dataset Version 2.0, where questions and answers are clearer and better organized, the F1-score improved to 0.4077. The highest results are obtained using Dataset Version 3.0, with concise and well-organized answers, resulting in an F1-score of 0.5258. However, the ROUGE-L score showed only a modest improvement, ranging from 0.2433 (Version 1.0) to 0.4144 (Version 3.0), indicating that the model still struggles with exact sequence alignment. On the other hand, the BERT F1 score shows the most significant improvement, moving from 0.5432 (Version 1.0) to 0.7281 (Version 3.0), highlighting the model’s ability to better capture semantic understanding as the data becomes more refined. While the results are improving with dataset refinement, introducing more data from diverse contexts can improve this performance more. 

% The other measures like ROUGE-L and BERT F1 also reflected the same trend where the best scores are achieved in Version 3.0.

% On the first test set, after fine-tuning with dataset Version 1.0, the F1-score does not improve significantly due to noise and unclear question-answer pairs. After fine-tuning with dataset Version 2.0, the F1-score improves as both questions and answers are clearer and more structured. The best performance is achieved with fine-tuning on Version 3.0 of the dataset, where answers are both clear and concise.

\begin{table}[h]
\vspace{-0.1in}
\caption{Evaluation metrics on test set after fine-tuning.}
\vspace{-0.1in}
\scriptsize % Reduce font size slightly for better fit
\centering
\begin{tabular}{|l|c|c|c|c|}
\hline
\textbf{Metric}      & \textbf{Base model} & \textbf{Version 1.0} & \textbf{Version 2.0} & \textbf{Version 3.0} \\ \hline
F1             & 0.3500              & 0.4280               & 0.4077               & 0.5258               \\ \hline
ROUGE-L        & 0.3073              & 0.2433               & 0.2440               & 0.4144               \\ \hline
BERT F1        & 0.6064              & 0.5432               & 0.5708               & 0.7281               \\ \hline
\end{tabular}

\label{tab:test_results}
\end{table}

\begin{figure}[h]
\centering
\includegraphics[width=0.45\textwidth, height=4cm]{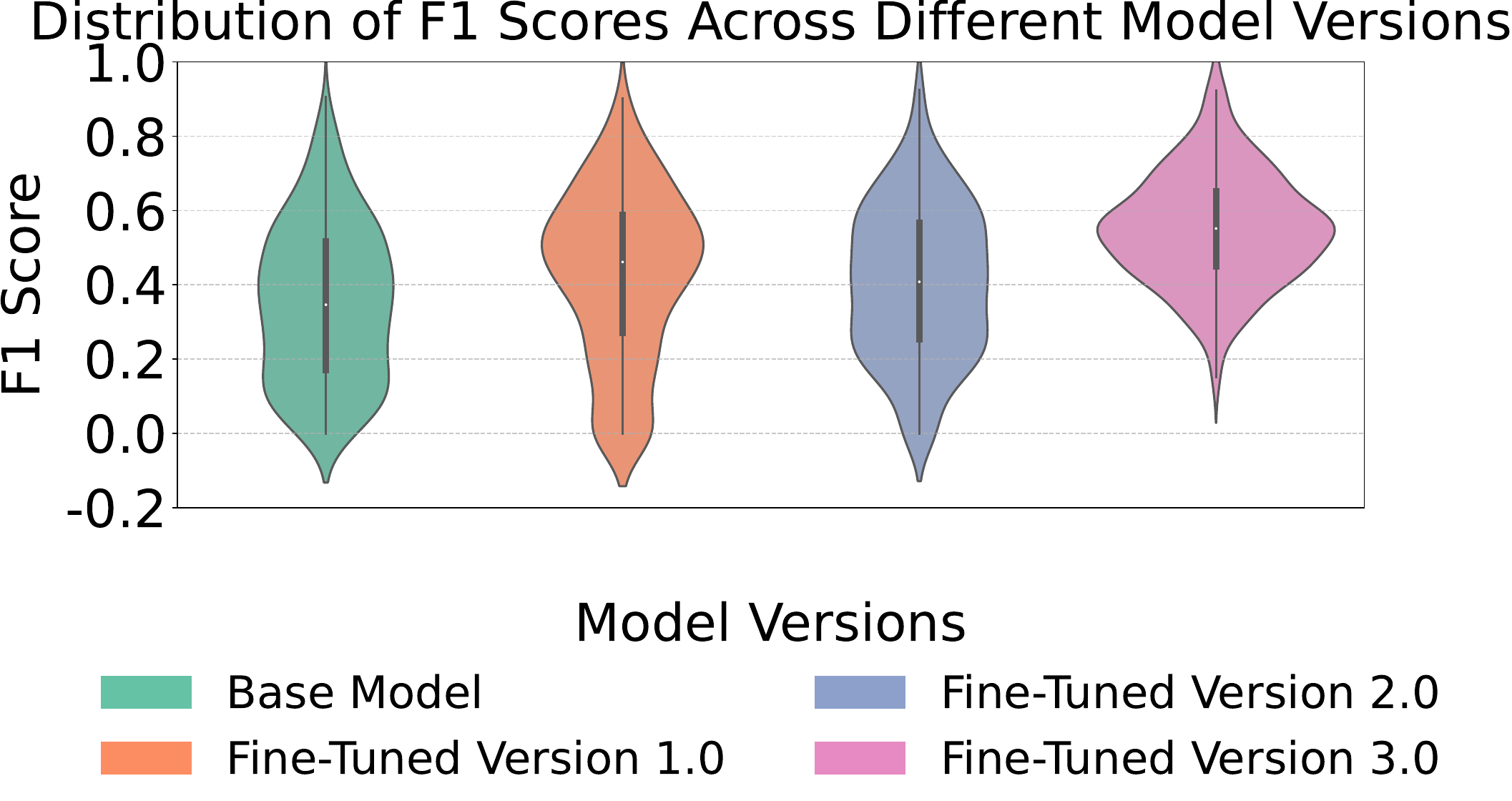}
\vspace{-0.1in}
\caption{Distribution of F1 scores on test set across different model versions.}
\vspace{-0.1in}
\label{fig:bertscore_distribution_F1}
\end{figure}

\begin{figure}[h]
\centering
\includegraphics[width=0.45\textwidth, height=4cm]{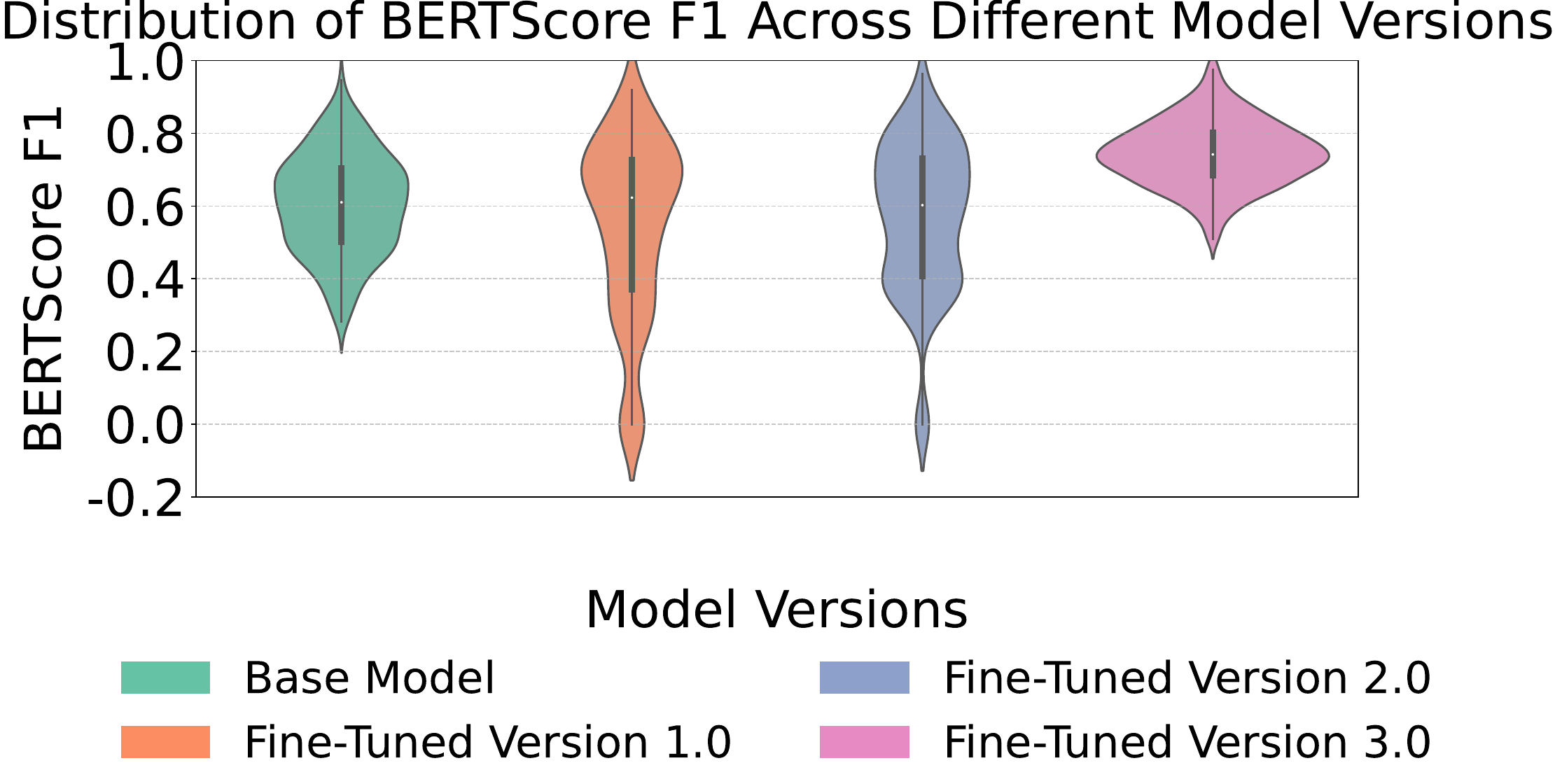}
\vspace{-0.1in}
\caption{Distribution of BERT F1 on test set across different models.}
\vspace{-0.1in}
\label{fig:bertscore_distribution_BertF1}
\end{figure}

% The results clearly demonstrate that our dataset refinement techniques (defined in Section \ref{sec: dataset_refinement}) significantly improve dataset quality, as evidenced by the performance on the test set across all three evaluation metrics: F1 Score, ROUGE-L Score, and BERT F1 Score. The results show enhanced outcomes compared to the base model and the fine-tuning with Version 1.0 of the dataset. The result underscores the need for cleaning up the datasets, as noise in the dataset hinders the model's ability to generate accurate answers. The refined version of the datasets facilitates much greater model performance in terms of domain-specific question-answering in particular. For instance, the F1 score increased from 0.3500 (Base Model) to 0.5258 (Version 3.0), reflecting the positive impact of clearer and better-organized data. Similarly, the BERT F1 score rose from 0.5432 (Version 1.0) to 0.7281 (Version 3.0), showing that the model benefits most from the improved dataset in capturing semantic understanding. The semi-automated refinement approach used in this study proves to be highly effective and adaptable, offering insights for improving dataset quality across various domains during the dataset construction process.  

The results clearly demonstrate that our dataset refinement techniques enhance dataset quality, as seen in the improved test set performance across all three evaluation metrics: F1 Score, ROUGE-L Score, and BERT F1 Score. Compared to the base model and fine-tuning with Version 1.0, the refined dataset leads to better model performance. For example, the F1 score increases from 0.3500 (Base Model) to 0.5258 (Version 3.0), and the BERT F1 score increases from 0.5432 (Version 1.0) to 0.7281 (Version 3.0). Figures \ref{fig:bertscore_distribution_F1} and \ref{fig:bertscore_distribution_BertF1} further support these findings by illustrating the distribution of F1 Scores  and BERT F1 Scores across different dataset versions. The violin plots show a clear upward shift in scores after fine-tuning with the refined datasets, with Version 3.0 achieving the highest concentration of high scores. This confirms that dataset refinement leads to more consistent and accurate model predictions, reducing performance variance and enhancing answer quality.

% \subsubsection*{Key Findings}

% Our semi-automated dataset refinement improves model performance, with Version 3.0 consistently achieving the best results, validating the effectiveness of our approach for domain-specific question answering. With more number of structured, diverse data, this performance can be improved. 

\subsection{Evaluating the Impact of 2-Stage Fine-Tuning}
% on T5 and Flan-T5 Model Variants}

In this section, we evaluate the impact of 2-stage fine-tuning on T5 and Flan-T5 variants using the Version 3.0 dataset to identify the maximum performance improvements achievable measured by F1 Score, ROUGE-L Score, and BERT F1 Score. Test set results are detailed in Table \ref{tab:merged_results}.

\begin{table}[ht]
\vspace{-0.1in}
\caption{Different models performance on test set (FT=Fine Tuning)}
\vspace{-0.1in}
\scriptsize
\centering
\begin{tabular}{|p{1.5cm}|p{1.5cm}|p{1cm}|p{1.3cm}|p{1.3cm}|}
\hline
\textbf{Model} & \textbf{Metric} & \textbf{Baseline} & \textbf{1-Stage FT} & \textbf{2-Stage FT} \\ \hline

\multirow{3}{*}{T5-small} 
& F1       & 0.3519 & 0.5584 & 0.5394 \\ 
& ROUGE-L  & 0.3094 & 0.4489 & 0.4250 \\ 
& BERT F1  & 0.6108 & 0.7491 & 0.7363 \\ \hline

\multirow{3}{*}{Flan-T5-small} 
& F1       & 0.1924 & 0.4894 & 0.4788 \\ 
& ROUGE-L  & 0.1667 & 0.3814 & 0.3703 \\ 
& BERT F1  & 0.4778 & 0.7067 & 0.6937 \\ \hline

\multirow{3}{*}{T5-base} 
& F1       & 0.3500 & 0.5258 & 0.4963 \\ 
& ROUGE-L  & 0.3073 & 0.4144 & 0.3928 \\ 
& BERT F1  & 0.6064 & 0.7281 & 0.7106 \\ \hline

\multirow{3}{*}{Flan-T5-base} 
& F1       & 0.1900 & 0.4937 & 0.4605 \\ 
& ROUGE-L  & 0.1701 & 0.3901 & 0.3524 \\ 
& BERT F1  & 0.4768 & 0.7074 & 0.6805 \\ \hline

\multirow{3}{*}{Flan-T5-large} 
& F1       & 0.2113 & 0.4900 & 0.4472 \\ 
& ROUGE-L  & 0.1887 & 0.3891 & 0.3347 \\ 
& BERT F1  & 0.4992 & 0.7036 & 0.6681 \\ \hline
\end{tabular}

\label{tab:merged_results}
\end{table}

After the first stage of fine-tuning, we observe that the model's performance improved as expected. Before fine-tuning, the F1 Score ranges between 0.19 to 0.35, but after fine-tuning, it has improved to a range of 0.45 to 0.55, with the largest increase observed in T5-small (from 0.3519 to 0.5584). However, after fine-tuning, most models achieve similar F1 scores, indicating that larger models do not yield significant additional improvements. Similarly, BERTScore F1 initially ranges between 0.47 to 0.61, increasing to 0.68 to 0.75 after fine-tuning, with T5-small again showing the highest gain. Despite these improvements, all models converges to a similar performance range, suggesting that fine-tuning benefits all model sizes but does not necessarily favor larger architectures.
Initially, the Flan-T5 models perform significantly worse than their T5 counterparts, as seen in the F1 Score and BERTScore F1, where Flan-T5-small (77M parameters) has an F1 of 0.1924 compared to 0.3519 for T5-Small (60M parameters), and Flan-T5-base (248M parameters) have an F1 of 0.1900 versus 0.3500 for T5-base (220M parameters). However, after fine-tuning, Flan models rapidly catch up to T5, with Flan-T5-small reaching 0.4894 F1, close to T5-small’s 0.5584, and Flan-T5-base achieving 0.4937 F1, nearly matching T5-base’s 0.5258. This trend is also evident in BERTScore F1, demonstrating that fine-tuning significantly enhances Flan models' ability to generate contextually aligned responses, despite their weaker initial performance.
% \textcolor{red}{On test set-2, the T5-small model surprisingly outperforms all others, despite having fewer parameters. Furthermore, T5 models slightly outperform FLAN models. This outcome can be attributed to the nature of the test set, which features short, precise, frequently asked queries, allowing the simpler T5-small architecture to excel in delivering accurate responses.}
After the second stage of fine-tuning with the synthetic dataset, contrary to our expectations, the model performance does not improve; in some cases, it even deteriorates. The primary reason for this outcome appears to be the synthetic data generation process. When synthetic questions are generated using the LLM, the model, due to its limited contextual understanding of the smart home domain, is unable to create sufficiently diverse questions. As a result, the synthetic dataset closely resembles the original dataset, leading to overfitting during fine-tuning. This lack of diversity limits the effectiveness of the synthetic dataset, causing the models to perform suboptimally. Our current generation method mainly rephrases existing QA pairs, leading to limited semantic diversity. This restricts generalization and may cause mild overfitting during fine-tuning. Therefore, we recommend choosing datasets based on validation performance. Fine-tuning on Version 3.0 alone typically gives strong, consistent results. Synthetic data can be added selectively to improve generalization for varied or unseen queries.

Despite limited data and computation, the final F1 and BERTScore F1 are satisfactory. We avoid larger models due to resource constraints, but the high BERTScore F1 shows the models still produce semantically accurate answers, validating our approach.

\begin{figure}[ht]
\centering
% \vspace{-0.1in}
\includegraphics[width=0.45\textwidth, height=4.5cm]{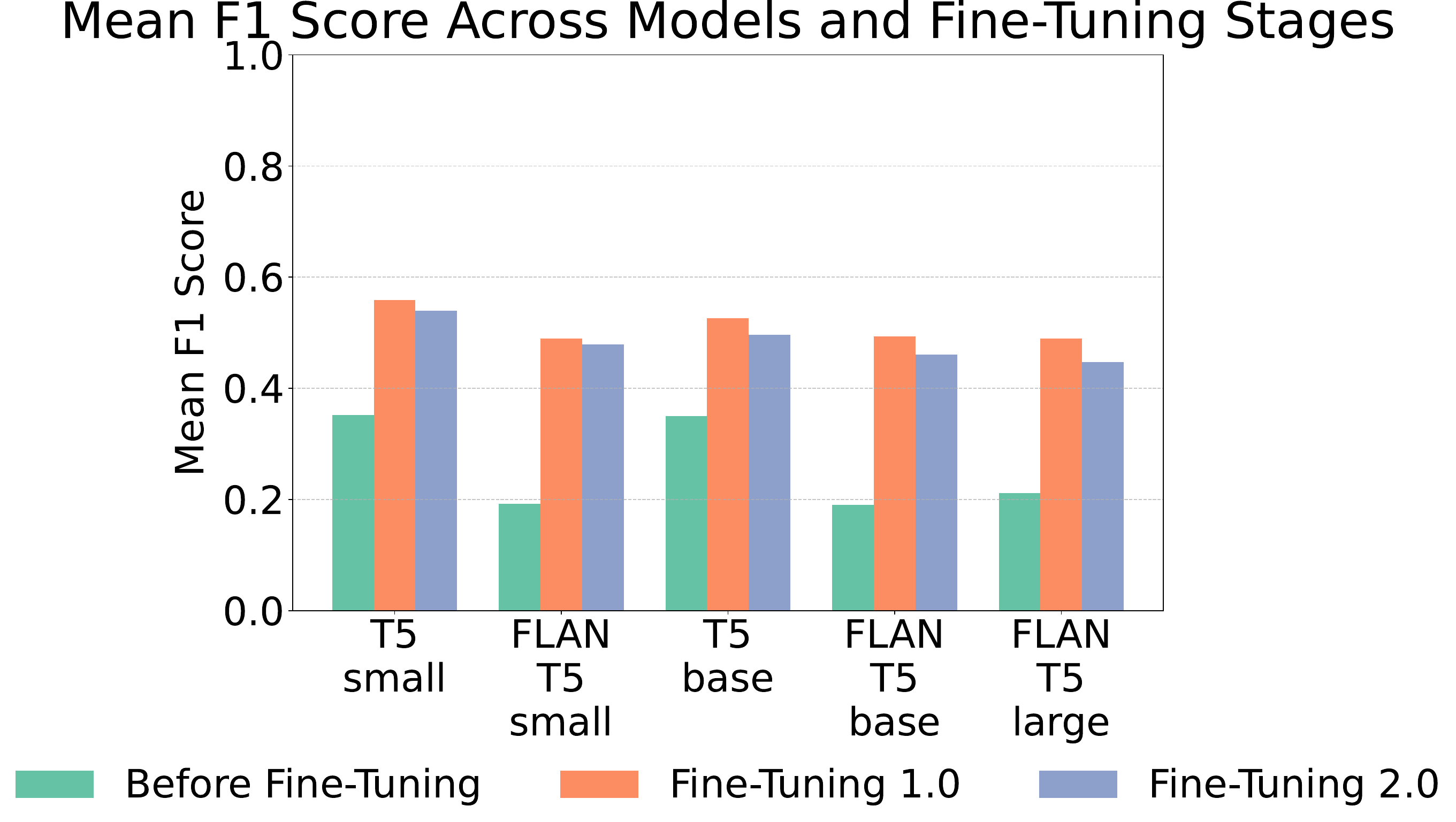}
\vspace{-0.2in}
\caption{Comparison of mean F1 Score on test set-1.}
\vspace{-0.1in}
\label{fig:mean_f1_score}
\end{figure}

\begin{figure}[ht]
\centering
% \vspace{-0.1in}
\includegraphics[width=0.45\textwidth, height=4.5cm]{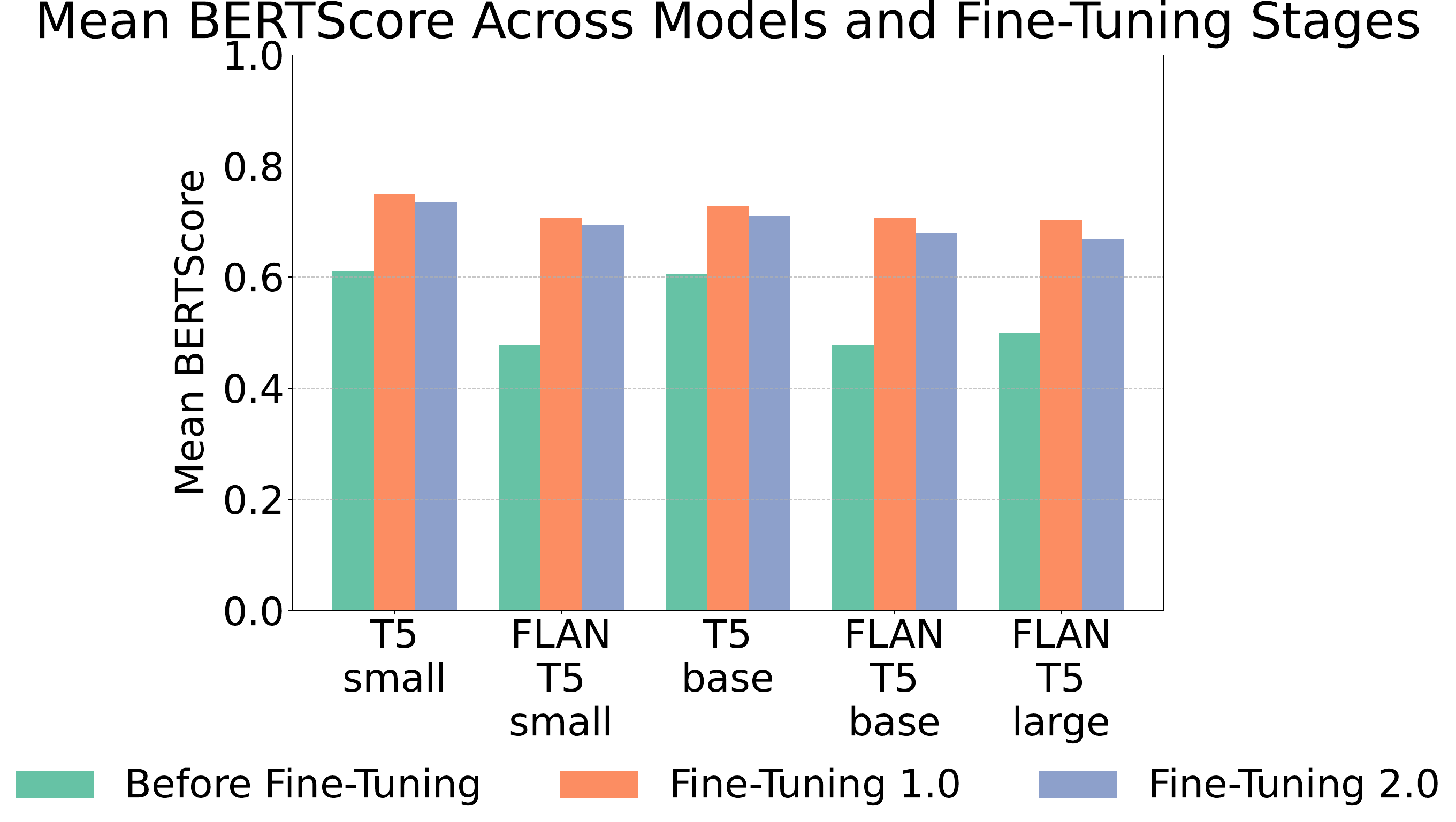}
\vspace{-0.1in}
\caption{Comparison of mean BERTScore F1 on test set-1.}
\vspace{-0.1in}
\label{fig:mean_bertscore}
\end{figure}

One noteworthy observation is that, after the first stage of fine-tuning, the five models performed similarly across all metrics, despite the differences in parameter counts. Furthermore, the bar plots in Figures \ref{fig:mean_f1_score} and \ref{fig:mean_bertscore} show that the average F1 and BERTScore F1 scores across the models exhibit no significant differences. Although the Flan-T5-large model required more computational resources, the performance gains are minimal. This suggests that for smaller datasets, larger models do not necessarily lead to better performance.

% After the second stage of fine-tuning, the performance of all models either stagnated or deteriorated, as shown in the bar plots. This decline can be attributed to the approach we used for synthetic data generation. Although the method aimed to diversify the dataset by generating similar questions, it did not introduce significant diversity. As a result, the models did not benefit from the additional fine-tuning, leading to performance stagnation.
% \begin{figure}[ht]
% \centering
% \begin{minipage}{0.47\textwidth}
%   \centering
%   \includegraphics[width=\linewidth,height=4.5cm]{mean_f1_score_histogram.pdf}
% \end{minipage}%
% \hspace{0.5cm}
% \begin{minipage}{0.47\textwidth}
%   \centering
%   \includegraphics[width=\linewidth,height=4.5cm]{mean_bertscore_histogram.pdf}
% \end{minipage}
% \caption{Comparison of mean F1 Score and mean BERTScore F1 on test set-1 across different models (T5-small, Flan-T5-small, T5-base, Flan-T5-base, Flan-T5-large) and fine-tuning stages.}
% \label{fig:avg_f1}
% \end{figure}

% \vspace{-0.5cm} % Adjust the negative space value as per your need

% \vspace{-0.5cm} % Adjust the negative space value as per your need

% \begin{tcolorbox}[colback=gray!20, colframe=black, sharp corners, boxrule=0.5pt, width=\textwidth]
% \textbf{Observation:} For our dataset, comparatively larger models yield minimal benefit despite increased computational resources.
% \end{tcolorbox}

% In conclusion, despite our limitations, we achieve notable improvements through fine-tuning for answering domain-specific questions from context. We observe that for fine-tuning models to answer questions from context, those with higher parameter counts do not exhibit significant performance gains compared to models with fewer parameters. Although the synthetic dataset does not enhance performance in this case, it highlights a promising approach: with the integration of more contextual data, LLMs can generate a more diverse synthetic dataset from the original, potentially aiding future synthetic dataset creation efforts significantly.

\subsection{Comparsion with Different Models}

% This section addresses RQ3 by summarizing the resource requirements and costs of the models evaluated as well as performance comparison, as shown in Table \ref{tab:performance_compare} and Table \ref{tab:resource_consume}. Open-source models (T5-Small, Llama-3.2-3B, and Llama-3.3-70B) offer flexibility without usage costs, while closed-source models (GPT-4.omini, GPT-4.o, and Gemini-1.5-pro) come with additional costs.

% The open-source models reveal a wide range of resource requirements. T5-Small with 223 million parameters is lightweight and requires just 1.07 GB of memory, an affordable option for low-resource environments. Llama-3.2-3B with 3 billion parameters requires 14.4 GB of memory. Llama-3.3-70B, on the other hand, with a much greater number of parameters at 70 billion, requires 336 GB of memory, which is problematic for hardware scalability and efficiency in operation. 

% Despite these differences in memory consumption, the performance of our fine-tuned model is greater than these models. As can be seen in Table \ref{tab:performance_compare}, our fine-tuned model achieves the best Bert F1 score among the Llama models. Thus, the T5-Base model stands out by offering greatly reduced resource usage as well as performance, and hence it is a more economically viable and practical tool for use in resource-constrained settings.

%version-02
We compare (in Tables \ref{tab:performance_compare_without_context}, \ref{tab:performance_compare_with_context} and \ref{tab:resource_consume}) our fine-tuned models with both open-source and closed-source LLMs to evaluate their effectiveness in domain-specific QA tasks. The results highlight a key trade-off: while large models like GPT-4o and Llama-3.3-70B perform well in zero-shot settings, they do not always maintain a significant advantage when domain-specific fine-tuning is applied to smaller models. 
% The results are detailed .

Examining the results without additional context in Table \ref{tab:performance_compare_without_context}, GPT-4o achieves the highest F1 score of 0.40 and a BERT F1 of 0.62, whereas the non-fine-tuned T5-base lags behind at 0.16 and 0.41, respectively. This is expected, as large proprietary models are pre-trained on vast datasets and can generalize well without needing additional adaptation. However, once fine-tuned on domain-specific data, T5-base closes the gap significantly, improving from an F1 score of 0.16 to 0.25 and from a BERT F1 of 0.41 to 0.54, outperforming Llama-3.2-3B and approaching Llama-3.3-70B’s performance. This suggests that while general models excel in zero-shot scenarios, proper fine-tuning allows smaller models to become competitive in specialized tasks.

When a context is provided in QA systems, the fine-tuned T5-base rivals GPT-4o, achieving an F1 score of 0.52 compared to GPT-4o’s 0.50, and a BERT F1 of 0.72, nearly matching GPT-4o’s 0.74 as shown in Table \ref{tab:performance_compare_with_context}. This highlights the effectiveness of fine-tuning, as a well-trained smaller model can leverage domain knowledge more efficiently than a much larger model that relies solely on general pre-trained information aligning with the findings in \cite{fu2024tiny}. Llama-3.3-70B, despite being significantly larger, does not surpass fine-tuned T5-base in F1 score, achieving only 0.45 compared to 0.52, indicating that increasing model size alone does not guarantee superior performance when fine-tuning is applied.

\begin{table}[ht]
\tiny  % Keep the font size small for better fit
\vspace{-0.1in}
\caption{Performance comparison of different models without additional context (zero-shot settings).}
\centering
\resizebox{\columnwidth}{!}{
\begin{tabular}{|c|c|c|c|c|c|c|}
\hline
\textbf{Metric} & \textbf{T5-base} & \textbf{Fine-tuned T5-base} &  \textbf{GPT-4omini} & \textbf{GPT-4o} & \textbf{Llama-3.3-70B} & \textbf{Llama-3.2-3B} \\ \hline
F1        & 0.16 & 0.25 & 0.36 & 0.40 & 0.33 & 0.30 \\ \hline
Rouge-L   & 0.16 & 0.24 & 0.23 & 0.26 & 0.22 & 0.20 \\ \hline
BERT F1   & 0.41 & 0.54 & 0.58 & 0.62 & 0.56 & 0.50 \\ \hline
\end{tabular}%
}
\label{tab:performance_compare_without_context}
\end{table}

% \begin{table}[ht]
% \tiny
% \vspace{-0.2in}
% \caption{Performance comparison of different models with context.}
% \vspace{-0.1in}
% \centering
% \resizebox{0.5\textwidth}{!}{
% \begin{tabular}{|c|c|c|c|c|c|c|c|}
% \hline
% \textbf{Metric} & \textbf{T5-base} & \textbf{Fine-tuned T5-base}  & \textbf{GPT-4omini} & \textbf{GPT-4o} & \textbf{Llama-3.3-70B} & \textbf{Llama-3.2-3B} \\ \hline
% {F1} & 0.35 & 0.52  & 0.49 & 0.50 & 0.45 & 0.43 \\ \hline
% {Rouge-L} & 0.30 & 0.41  & 0.33 & 0.36 & 0.32 & 0.33 \\ \hline
% {BERT F1} & 0.60 & 0.72  & 0.69 & 0.74 & 0.65 & 0.63 \\ \hline
% \end{tabular}%
% }
% \label{tab:performance_compare_with_context}
% \end{table}

\begin{table}[ht]
\tiny
\vspace{-0.2in}
\caption{Performance comparison of different models with context.}
\vspace{-0.1in}
\centering
\resizebox{\columnwidth}{!}{
\begin{tabular}{|c|c|c|c|c|c|c|}
\hline
\textbf{Metric} & \textbf{T5-base} & \textbf{Fine-tuned T5-base}  & \textbf{GPT-4omini} & \textbf{GPT-4o} & \textbf{Llama-3.3-70B} & \textbf{Llama-3.2-3B} \\ \hline
{F1} & 0.35 & 0.52  & 0.49 & 0.50 & 0.45 & 0.43 \\ \hline
{Rouge-L} & 0.30 & 0.41  & 0.33 & 0.36 & 0.32 & 0.33 \\ \hline
{BERT F1} & 0.60 & 0.72  & 0.69 & 0.74 & 0.65 & 0.63 \\ \hline
\end{tabular}%
}
\label{tab:performance_compare_with_context}
\end{table}

Besides, the resource consumption table further reinforces the practicality of lightweight models. T5-base requires just 1.07GB of memory, making it drastically more efficient than Llama-3.3-70B, which demands 336GB, and even Llama-3.2-3B, which requires 14.4GB. This efficiency makes fine-tuned models much easier to deploy in domain-specific applications, especially where computational resources are limited. Moreover, proprietary models like GPT-4o come with recurring API costs, making them financially impractical for large-scale, long-term use, whereas T5-base and Llama variants can be deployed cost-free as open-source models.

\begin{table}[ht]
\vspace{-0.1in}
\caption{Resource consumption for each model.}
\tiny
\resizebox{\columnwidth}{!}{%
\begin{tabular}{|l|l|l|l|l|l|}
\hline
\textbf{}                 & \textbf{T5-base} & \textbf{GPT-4omini} & \textbf{GPT-4o} & \textbf{Llama-3.3-70B} & \textbf{Llama-3.2-3B} \\ \hline
{Size}             & 223M             & NA                  & NA              & 70B                    & 3B                     \\ \hline
{\begin{tabular}[c]{@{}l@{}}Open Source\\ or Not\end{tabular}} & Open Source       & Closed Source        & Closed Source  & Open Source            & Open Source            \\ \hline
{Resource Needed}  & 1.07GB           & NA                  & NA              & 336GB                  & 14.4GB                 \\ \hline
{Usage Cost}       & No               & Yes                 & Yes             & No                      & No                     \\ \hline
\end{tabular}%
}
\label{tab:resource_consume}
\end{table}

These findings demonstrate that while large pre-trained models have an initial advantage in zero-shot settings, smaller models, when fine-tuned correctly, can rival or even surpass them in domain-specific applications. Instead of relying on ever-larger models, optimizing smaller models through fine-tuning and leveraging context appears to be a more effective and resource-efficient strategy.

\section{Conclusion}

In this work, we aimed to investigate the extent of improvements that can be achieved in smaller LLMs for answering questions from context within the smart home security domain by fine-tuning them with our novel dataset. During this process, we closely examined the impact of dataset quality on model performance. Additionally, we explored the potential of synthetic data generated by large language models (LLMs) to enhance the results, identifying key factors that influence model effectiveness. This work not only highlights the potential but also addresses the challenges of using smaller LLMs in the field of smart home IoT security and privacy, laying a solid foundation for future advancements.
Key limitations include the scalability of the context generation pipeline—currently reliant on LLMs and manual review—which hampers automation. Incorporating information retrieval (IR) could improve scalability by dynamically sourcing relevant content, enabling a more efficient and adaptive IR+LLM system for real-world deployment. Additionally, the lack of user studies limits understanding of practical usability, marking a crucial area for future work. Enhancing synthetic data diversity through varied prompts, contrastive examples, or human input could also improve semantic richness and model robustness.
\section*{Acknowledgment}
The authors thank the anonymous reviewers for their valuable comments. This work is partially supported by the Natural Sciences and Engineering Research Council of Canada and the Department of National Defence Canada under the Discovery Grants RGPIN-2021-04106 and DGDND-2021-04106.

\bibliographystyle{IEEEtran}

\bibliography{references}

% Generated by IEEEtran.bst, version: 1.14 (2015/08/26)
\begin{thebibliography}{10}
\providecommand{\url}[1]{#1}
\csname url@samestyle\endcsname
\providecommand{\newblock}{\relax}
\providecommand{\bibinfo}[2]{#2}
\providecommand{\BIBentrySTDinterwordspacing}{\spaceskip=0pt\relax}
\providecommand{\BIBentryALTinterwordstretchfactor}{4}
\providecommand{\BIBentryALTinterwordspacing}{\spaceskip=\fontdimen2\font plus
\BIBentryALTinterwordstretchfactor\fontdimen3\font minus \fontdimen4\font\relax}
\providecommand{\BIBforeignlanguage}[2]{{%
\expandafter\ifx\csname l@#1\endcsname\relax
\typeout{** WARNING: IEEEtran.bst: No hyphenation pattern has been}%
\typeout{** loaded for the language `#1'. Using the pattern for}%
\typeout{** the default language instead.}%
\else
\language=\csname l@#1\endcsname
\fi
#2}}
\providecommand{\BIBdecl}{\relax}
\BIBdecl

\bibitem{vardakis2024review}
G.~Vardakis, G.~Hatzivasilis, E.~Koutsaki, and N.~Papadakis, ``Review of smart-home security using the internet of things,'' \emph{Electronics}, vol.~13, no.~16, p. 3343, 2024.

\bibitem{sahu2024exploring}
S.~K. Sahu and K.~Mazumdar, ``Exploring security threats and solutions techniques for internet of things ({IoT}): from vulnerabilities to vigilance,'' \emph{Frontiers in Artificial Intelligence}, vol.~7, p. 1397480, 2024.

\bibitem{vetrivel2023examining}
S.~Vetrivel, V.~Van~Harten, C.~H. Ga{\~n}{\'a}n, M.~Van~Eeten, and S.~Parkin, ``Examining consumer reviews to understand security and privacy issues in the market of smart home devices,'' in \emph{USENIX Security}, 2023.

\bibitem{deng2024pentestgpt}
G.~Deng, Y.~Liu, V.~Mayoral-Vilches, P.~Liu, Y.~Li, Y.~Xu, T.~Zhang, Y.~Liu, M.~Pinzger, and S.~Rass, ``{PentestGPT}: Evaluating and harnessing large language models for automated penetration testing,'' in \emph{USENIX Security}, 2024.

\bibitem{fang2024large}
C.~Fang, N.~Miao, S.~Srivastav, J.~Liu, R.~Zhang, R.~Fang, R.~Tsang, N.~Nazari, H.~Wang, H.~Homayoun \emph{et~al.}, ``Large language models for code analysis: Do {LLMs} really do their job?'' in \emph{USENIX Security}, 2024.

\bibitem{liu2024cyberbench}
Z.~Liu, J.~Shi, and J.~F. Buford, ``Cyberbench: A multi-task benchmark for evaluating large language models in cybersecurity,'' in \emph{AICS}, 2024.

\bibitem{chatiot}
S.~C. Ye~Dong, Yan Lin~Aung and J.~Zhou, ``{ChatIoT}: Large language model-based security assistant for internet of things with retrieval-augmented generation,'' \emph{arXiv preprint arXiv:2502.09896}, 2025.

\bibitem{aiot}
H.~Zhu, P.~Tiwari, A.~Ghoneim, and M.~S. Hossain, ``A collaborative ai-enabled pretrained language model for aiot domain question answering,'' \emph{IEEE Transactions on Industrial Informatics}, vol.~18, no.~5, pp. 3387--3396, 2021.

\bibitem{zeng2017end}
E.~Zeng, S.~Mare, and F.~Roesner, ``End user security and privacy concerns with smart homes,'' in \emph{SOUPS}, 2017.

\bibitem{touqeer2021smart}
H.~Touqeer, S.~Zaman, R.~Amin, M.~Hussain, F.~Al-Turjman, and M.~Bilal, ``Smart home security: challenges, issues and solutions at different {IoT} layers,'' \emph{The Journal of Supercomputing}, vol.~77, no.~12, pp. 14\,053--14\,089, 2021.

\bibitem{chung2024scaling}
H.~W. Chung, L.~Hou, S.~Longpre, B.~Zoph, Y.~Tay, W.~Fedus, Y.~Li, X.~Wang, M.~Dehghani, S.~Brahma \emph{et~al.}, ``Scaling instruction-finetuned language models,'' \emph{Journal of Machine Learning Research}, vol.~25, no.~70, pp. 1--53, 2024.

\bibitem{hallucinate1}
L.~Huang, W.~Yu, W.~Ma, W.~Zhong, Z.~Feng, H.~Wang, Q.~Chen, W.~Peng, X.~Feng, B.~Qin \emph{et~al.}, ``A survey on hallucination in large language models: Principles, taxonomy, challenges, and open questions,'' \emph{ACM Transactions on Information Systems}, 2024.

\bibitem{google_nest_community}
\BIBentryALTinterwordspacing
{Google Nest Community}. (2024) Google nest community forum. Accessed: 2024-09-14. [Online]. Available: \url{https://www.googlenestcommunity.com}
\BIBentrySTDinterwordspacing

\bibitem{apple_community_discussions}
\BIBentryALTinterwordspacing
{Apple}. (2024) Apple community discussions forum. Accessed: 2024-09-14. [Online]. Available: \url{https://discussions.apple.com/}
\BIBentrySTDinterwordspacing

\bibitem{verizon_community_forums}
\BIBentryALTinterwordspacing
{Verizon Community Forums}. (2024) Verizon community forums. Accessed: 2024-09-14. [Online]. Available: \url{https://community.verizon.com/t5/forums}
\BIBentrySTDinterwordspacing

\bibitem{blei2003latent}
D.~M. Blei, A.~Y. Ng, and M.~I. Jordan, ``Latent dirichlet allocation,'' \emph{Journal of machine Learning research}, vol.~3, pp. 993--1022, 2003.

\bibitem{guhr2020privacy}
N.~Guhr, O.~Werth, P.~P.~H. Blacha, and M.~H. Breitner, ``Privacy concerns in the smart home context,'' \emph{SN Applied Sciences}, vol.~2, pp. 1--12, 2020.

\bibitem{zhong2020jec}
H.~Zhong, C.~Xiao, C.~Tu, T.~Zhang, Z.~Liu, and M.~Sun, ``{JEC-QA}: a legal-domain question answering dataset,'' in \emph{AAAI}, 2020.

\bibitem{yadav2022chq}
S.~Yadav, D.~Gupta, and D.~Demner-Fushman, ``{CHQ-Summ}: A dataset for consumer healthcare question summarization,'' \emph{arXiv preprint arXiv:2206.06581}, 2022.

\bibitem{tihanyi2024cybermetric}
N.~Tihanyi, M.~A. Ferrag, R.~Jain, and M.~Debbah, ``{CyberMetric}: A benchmark dataset for evaluating large language models knowledge in cybersecurity,'' \emph{arXiv preprint arXiv:2402.07688}, 2024.

\bibitem{aghaei2023securebert}
E.~Aghaei, X.~Niu, W.~Shadid, and E.~Al-Shaer, ``{SecureBERT}: A domain-specific language model for cybersecurity,'' in \emph{SecureComm}, 2023.

\bibitem{liu2023secqa}
Z.~Liu, ``{SecQA}: A concise question-answering dataset for evaluating large language models in computer security,'' \emph{arXiv preprint arXiv:2312.15838}, 2023.

\bibitem{wmdp}
N.~Li, A.~Pan, A.~Gopal, S.~Yue, D.~Berrios, A.~Gatti, J.~D. Li, A.-K. Dombrowski, S.~Goel, G.~Mukobi \emph{et~al.}, ``The {WMDP} benchmark: measuring and reducing malicious use with unlearning,'' in \emph{ICML}, 2024.

\bibitem{avs_forum}
\BIBentryALTinterwordspacing
{AVS Forum}, ``{AVS Forum - Home Theater Discussions and Reviews},'' 2024, accessed: 2024-09-30. [Online]. Available: \url{https://www.avsforum.com/forums/}
\BIBentrySTDinterwordspacing

\bibitem{diychatroom_forum}
\BIBentryALTinterwordspacing
{DIY}, ``{DIY Chatroom - Home Improvement Forum},'' 2024, accessed: 2024-09-30. [Online]. Available: \url{https://www.diychatroom.com}
\BIBentrySTDinterwordspacing

\bibitem{cocoontech_forum}
\BIBentryALTinterwordspacing
{CocoonTech Forum}, ``{CocoonTech - Home Automation and Security Forum},'' 2024, accessed: 2024-09-30. [Online]. Available: \url{https://cocoontech.com/forum/}
\BIBentrySTDinterwordspacing

\bibitem{digitalhome_forum}
\BIBentryALTinterwordspacing
{Digital Home Forum}, ``{Digital Home - Canadian Digital Technology Discussion},'' 2024, accessed: 2024-09-30. [Online]. Available: \url{https://www.digitalhome.ca/forums/}
\BIBentrySTDinterwordspacing

\bibitem{diynot_forum}
\BIBentryALTinterwordspacing
{DIYnot Forum}, ``{DIYnot - Home Improvement and DIY Advice},'' 2024, accessed: 2024-09-30. [Online]. Available: \url{https://www.diynot.com/diy/}
\BIBentrySTDinterwordspacing

\bibitem{ezlo_forum}
\BIBentryALTinterwordspacing
{Ezlo Community}, ``{Ezlo Community - Smart Home Automation and Security},'' 2024, accessed: 2024-09-30. [Online]. Available: \url{https://community.ezlo.com}
\BIBentrySTDinterwordspacing

\bibitem{home_assistant_forum}
\BIBentryALTinterwordspacing
{Home Assistant Community}, ``{Home Assistant Community - Open Source Home Automation},'' 2024, accessed: 2024-09-30. [Online]. Available: \url{https://community.home-assistant.io}
\BIBentrySTDinterwordspacing

\bibitem{reddit_forum}
\BIBentryALTinterwordspacing
{Reddit}, ``{Reddit - Online Discussion and Community Platform},'' 2024, accessed: 2024-09-30. [Online]. Available: \url{https://www.reddit.com}
\BIBentrySTDinterwordspacing

\bibitem{smartthings_forum}
\BIBentryALTinterwordspacing
{SmartThings Community}, ``{SmartThings Community - Home Automation Discussions},'' 2024, accessed: 2024-09-30. [Online]. Available: \url{https://community.smartthings.com}
\BIBentrySTDinterwordspacing

\bibitem{snb_forum}
\BIBentryALTinterwordspacing
{SmallNetBuilder Forums}, ``{SNB Forums - Networking, IoT, and Smart Home Discussions},'' 2024, accessed: 2024-09-30. [Online]. Available: \url{https://www.snbforums.com}
\BIBentrySTDinterwordspacing

\bibitem{stackexchange_iot}
\BIBentryALTinterwordspacing
{Stack Exchange}, ``{Internet of Things Stack Exchange - Q\&A for IoT Enthusiasts},'' 2024, accessed: 2024-09-30. [Online]. Available: \url{https://iot.stackexchange.com}
\BIBentrySTDinterwordspacing

\bibitem{openwrt_forum}
O.~Forum, ``Openwrt forum - open source router firmware and networking discussions,'' \url{https://forum.openwrt.org}, 2024, accessed: 2025-01-16.

\bibitem{level1techs_forum}
\BIBentryALTinterwordspacing
{Level1Techs Forum}, ``{Level1Techs Forum - Technology, Networking, and Security Discussions},'' 2024, accessed: 2024-09-30. [Online]. Available: \url{https://forum.level1techs.com}
\BIBentrySTDinterwordspacing

\bibitem{feedspot_automation_forum}
\BIBentryALTinterwordspacing
{Feedspot}. (2024) Home automation forums. Accessed: 2024-09-14. [Online]. Available: \url{https://forums.feedspot.com/home_automation_forums/}
\BIBentrySTDinterwordspacing

\bibitem{whirlpool_forum}
\BIBentryALTinterwordspacing
{Whirlpool Forums}. (2024) {Whirlpool - Australian Broadband News \& Forums}. Accessed: 2024-02-28. [Online]. Available: \url{https://forums.whirlpool.net.au/}
\BIBentrySTDinterwordspacing

\bibitem{quora_forum}
\BIBentryALTinterwordspacing
{Quora}. (2024) {Quora - Questions and Answers Community}. Accessed: 2024-02-28. [Online]. Available: \url{https://www.quora.com/}
\BIBentrySTDinterwordspacing

\bibitem{parsehub}
``Parsehub | free web scraping - the most powerful web scraper,'' \url{https://www.parsehub.com/}, 2023, accessed: 2025-04-16.

\bibitem{octoparse}
``Octoparse: Web scraping tool \& free web crawlers,'' \url{https://www.octoparse.com/}, 2016, accessed: 2025-02-16.

\bibitem{Google_AI}
G.~AI, ``Gemini developer api | gemma open models | google ai for developers,'' \url{https://ai.google.dev/}, 2023, accessed: 2025-04-16.

\bibitem{keklik2019rule}
O.~Keklik, T.~Tuglular, and S.~Tekir, ``Rule-based automatic question generation using semantic role labeling,'' \emph{IEICE TRANSACTIONS on Information and Systems}, vol. 102, no.~7, pp. 1362--1373, 2019.

\bibitem{fabbri2020template}
A.~R. Fabbri, P.~Ng, Z.~Wang, R.~Nallapati, and B.~Xiang, ``Template-based question generation from retrieved sentences for improved unsupervised question answering,'' \emph{arXiv preprint arXiv:2004.11892}, 2020.

\bibitem{kabongo2024effective}
S.~Kabongo, J.~D’Souza, and o.~Auer, ``Effective context selection in llm-based leaderboard generation: An empirical study,'' in \emph{NLDB}, 2024.

\bibitem{huggingface}
H.~Face, ``Hugging face – the ai community building the future,'' \url{https://huggingface.co/}, 2016, accessed: 2025-04-16.

\bibitem{LangChain}
LangChain, ``Langchain,'' \url{https://www.langchain.com/}, 2022, accessed: 2025-04-16.

\bibitem{alam2024ctibench}
M.~T. Alam, D.~Bhusal, L.~Nguyen, and N.~Rastogi, ``{CTIBench}: A benchmark for evaluating {LLMs} in cyber threat intelligence,'' \emph{arXiv preprint arXiv:2406.07599}, 2024.

\bibitem{rajpurkar2016squad}
P.~Rajpurkar, J.~Zhang, K.~Lopyrev, and P.~Liang, ``Squad: 100,000+ questions for machine comprehension of text,'' in \emph{EMNL}, 2016.

\bibitem{lin2004rouge}
C.-Y. Lin, ``Rouge: A package for automatic evaluation of summaries,'' in \emph{Text summarization branches out}, 2004, pp. 74--81.

\bibitem{zhang2019bertscore}
T.~Zhang, V.~Kishore, F.~Wu, K.~Q. Weinberger, and Y.~Artzi, ``{BERTScore}: Evaluating text generation with {BERT},'' \emph{arXiv preprint arXiv:1904.09675}, 2019.

\bibitem{fu2024tiny}
X.-Y. Fu, M.~T.~R. Laskar, E.~Khasanova, C.~Chen, and S.~B. TN, ``Tiny titans: Can smaller large language models punch above their weight in the real world for meeting summarization?'' \emph{arXiv preprint arXiv:2402.00841}, 2024.

\end{thebibliography}

\end{document}